\documentclass[aps,preprintnumbers,showpacs,twocolumn,superscriptaddress,floatfix,nofootinbib]{revtex4}

\usepackage{amsmath}
\usepackage{amssymb}
\usepackage{color}
\usepackage{dcolumn}
\usepackage{graphicx}
\usepackage{epstopdf}           
\usepackage{latexsym}
\usepackage{times}
\usepackage{ifpdf}

\newcommand{\bea}{\begin{eqnarray}}
\newcommand{\eea}{\end{eqnarray}}
\newcommand{\beq}{\begin{equation}}
\newcommand{\eeq}{\end{equation}}

\newcommand{\cf}{\textit{cf.}~}
\newcommand{\ie}{\textit{i.e.,}~}
\newcommand{\eg}{\textit{e.g.,}~}

\newcommand{\mpc}{{\rm Mpc}}

\newcommand{\pls}{~~~}

\def\scri{\mathcal{J}^+}

\begin{document}

\title{Gravitational-wave detectability of equal-mass black-hole
  binaries with aligned spins}

\author{Christian Reisswig}
\affiliation{
  Max-Planck-Institut f\"ur Gravitationsphysik,
  Albert-Einstein-Institut,
  Potsdam-Golm, Germany
}

\author{Sascha Husa}
\affiliation{Departament de F\'isica, Universitat de les Illes
  Balears, Palma de Mallorca, Spain
}
\affiliation{
  Max-Planck-Institut f\"ur Gravitationsphysik,
  Albert-Einstein-Institut,
  Potsdam-Golm, Germany
}

\author{Luciano Rezzolla}
\affiliation{
  Max-Planck-Institut f\"ur Gravitationsphysik,
  Albert-Einstein-Institut,
  Potsdam-Golm, Germany
}
\affiliation{
  Department of Physics and Astronomy,
  Louisiana State University,
  Baton Rouge, LA, USA
}

\author{Ernst Nils Dorband}
\affiliation{
  Max-Planck-Institut f\"ur Gravitationsphysik,
  Albert-Einstein-Institut,
  Potsdam-Golm, Germany
}

\author{Denis Pollney}
\affiliation{
  Max-Planck-Institut f\"ur Gravitationsphysik,
  Albert-Einstein-Institut,
  Potsdam-Golm, Germany
}

\author{Jennifer Seiler}
\affiliation{
  Max-Planck-Institut f\"ur Gravitationsphysik,
  Albert-Einstein-Institut,
  Potsdam-Golm, Germany
}

\date{\today}


\begin{abstract}
  Binary black-hole systems with spins aligned or anti-aligned to
  the orbital angular momentum, and which therefore do not exhibit
  precession effects, provide the natural ground to start detailed
  studies of the influence of strong-field spin effects on
  gravitational wave observations of coalescing binaries.
  Furthermore, such systems may be the preferred end-state of the
  inspiral of generic supermassive binary black-hole systems. In view
  of this, we have computed the inspiral and merger of a large set of
  binary systems of equal-mass black holes with spins parallel to the
  orbital angular momentum but otherwise arbitrary. Our attention is
  particularly focused on the gravitational-wave emission so as to
  quantify how much spin effects contribute to the signal-to-noise
  ratio, to the horizon distances, and to the relative event rates for
  the representative ranges in masses and detectors. As expected, the
  signal-to-noise ratio increases with the projection of the total
  black hole spin in the direction of the orbital momentum. We find
  that equal-spin binaries with maximum spin aligned with the orbital
  angular momentum are more than ``three times as loud'' as the
  corresponding binaries with anti-aligned spins, thus corresponding
  to event rates up to $30$ times larger. We also consider the
  waveform mismatch between the different spinning configurations and
  find that, within our numerical accuracy, binaries with opposite
  spins \mbox{$\boldsymbol{S}_1=-\boldsymbol{S}_2$} cannot be
  distinguished whereas binaries with spin
  $\boldsymbol{S}_1=\boldsymbol{S}_2$ have clearly distinct
  gravitational-wave emissions. Finally, we derive a simple expression
  for the energy radiated in gravitational waves and find that the
  binaries always have efficiencies $E_{\rm rad}/M \gtrsim 3.6\%$,
  which can become as large as $E_{\rm rad}/M \simeq 10\%$ for
  maximally spinning binaries with spins aligned with the orbital
  angular momentum. These binaries are therefore among the most
  efficient sources of energy in the Universe.
 \end{abstract}

\pacs{04.25.Dm, 04.30.Db, 95.30.Sf, 97.60.Lf} 
\maketitle

\section{Introduction} 
\label{sec:intro}

It has been a long-standing goal of the field of numerical relativity
to provide results for gravitational-wave data analysis and
thus enhance the capabilities of current and future gravitational wave
detectors, in particular regarding the observation of compact binary
coalescence. With a series of breakthroughs in 2005
\cite{Pretorius:2005gq,Campanelli:2005dd,Baker:2005vv}, this long-term
goal has suddenly become reality. However, much further work is
required to actually understand the practical implications of
numerical solutions of the full Einstein equations for
gravitational-wave data analysis.  Indeed, first studies suggest that
template banks that use numerical information can increase the reach
of detectors~\cite{Ajith:2007qp,Ajith:2007kx,Ajith:2007xh}, aid the
calibration of search pipelines
\cite{Aylott:2009ya,Farr:2009pg,Santamaria:2009tm}, and improve the
estimation of parameters, such as \eg sky
location~\cite{Babak:2008bu}.

In this paper we use gravitational waveforms from numerical-relativity
(NR) calculations of a number of sequences of equal-mass spinning
black-hole binaries whose spins are aligned (anti-aligned) with the
orbital angular momentum, and consider the detectability of these
binaries for the ground-based gravitational wave-detectors as well as
for the planned space-based LISA interferometer.

Our interest in this type of binary stems from the fact that there are
indications they represent preferred configurations in nature, at
least if the black holes are supermassive. It has been shown, in fact,
that when the binary is surrounded by a massive circumbinary disc, as
the one expected by the merger of two galaxies, the dissipative
dynamics of the matter produces a torque with the effect of aligning
the spins to the orbital angular momentum~\cite{Bogdanovic:2007hp}. In
addition, the merger of binaries with aligned spins yields recoil
velocities which are sufficiently small (\ie $\lesssim 450\,{\rm
  km/s}$~\cite{Koppitz-etal-2007aa,Herrmann:2007ex,Herrmann:2007ac})
to prevent the final black hole from being expelled from the host
galaxy. This would then be compatible with the overwhelming
astronomical evidence that massive black holes reside at the centers
of most galaxies.

Our parameter space is therefore 2-dimensional, parametrized by the
projections $a_1$, $a_2$ of the dimensionless spins ${\boldsymbol
  a}_i\equiv{\boldsymbol S}_i/M_i^2$ of the individual black holes on
to direction of the angular momentum (chosen as the $z$-axis).  As a
result, spins that are aligned with the orbital angular momentum are
characterized by positive values of $a_1$, $a_2$, while anti-aligned
spins have negative values. Previous studies of this parameter
space~\cite{Campanelli:2006uy, Campanelli:2006vp, Campanelli:2006gf,
  Buonanno:07b, Koppitz-etal-2007aa, Herrmann:2007ex, Herrmann:2007ac,
  Pollney:2007ss, Rezzolla-etal-2007, BoyleKesdenNissanke:07,
  BoyleKesden:07, Marronetti:2007wz,Rezzolla-etal-2007b,
  Rezzolla-etal-2007c,Barausse:2009uz}, have considered the recoil
velocity and final spin of the merger remnant, and have constructed
phenomenological formulas for these quantities given the initial spins
$a_1$ and $a_2$ of the binary.

In this work, we move our focus to the detectability of a given set of
binaries in the parameter sub-space of (anti-) aligned spins, \ie for
each of these binaries and across a set of different masses we
calculate the signal-to-noise ratio (SNR) for the
LIGO~\cite{Abbott:2007kv,Waldman06}, enhanced LIGO
(eLIGO)~\cite{Adhikari06}, advanced LIGO
(AdLIGO)~\cite{AdvLIGO,Ajith:2007kx}, Virgo~\cite{Acernese2006},
advanced Virgo (AdVirgo)~\cite{AdVirgo}, and
LISA~\cite{LISA1,Danzmann:2003tv} detectors .

In this way we attempt to address the following questions:
\begin{itemize}
\item[]\textit{(i)} Which among the aligned-spin configurations is the
  ``loudest'' and which one is the ``quietest''?

\item[]\textit{(ii)} How large is the difference in signal-to-noise
  ratio between the loudest and the quietest?

\item[]\textit{(iii)} How do these considerations depend on the detector
  used, the mass of the binary, and the number of harmonics?

\item[] \textit{(iv)} Are there configurations whose waveforms are
  difficult to distinguish and are hence degenerate in the space of
  templates?
\end{itemize}

Overall, and as expected, we find that equal-spinning, maximally
anti-aligned binaries generally produce the lowest SNR while
equal-spinning, maximally aligned binaries produce the highest SNR.
For any mass, the SNR can be well described with a low-order
polynomial of the initial spins $\rho=\rho(a_1, a_2)$ and generally it
increases with the total dimensionless spin along the angular momentum
direction, $a \equiv \frac{1}{2}(\boldsymbol{a}_1 + \boldsymbol{a}_2
)\cdot \boldsymbol{\hat{L}}$. The possibility of describing the whole
behaviour of the waveforms from equal-mass, aligned/antialigned
binaries in terms of a single scalar quantity, namely $a$, provides a
certain amount of optimism that also more complex spin configurations
can, ultimately, be described in terms of a few parameters only.

We also analyze the impact that modes of the gravitational-wave field
of order larger than $\ell=2$ but smaller than $\ell=5$ have on the
maximum SNR and show that for low masses $M\in[20,100]$ they
contribute, say for the LIGO detector, $\approx 2.5\%$, whereas for
intermediate masses $M>100\,\,M_{\odot}$ they contribute $\approx
8\%$~\footnote{Note that for some specific angles at which the SNR is
  not maximum, the contribution of the higher modes can be much more
  significant}. In addition, we determine the ratio between maximum
and averaged SNR for $\ell>2$ which is known to be $\sqrt{5}$ when
considering only the $\ell=2, m=2$ mode. We also calculate the
mismatch between the waveforms from different binaries across our
spin-diagram and find that binaries along the diagonal $a_1=-a_2$
cannot be distinguished within our given numerical accuracy, whereas
configurations along the diagonal $a_1=a_2$ are clearly different (\cf
Fig~\ref{fig:match_1} and~\ref{fig:match_2}, as well as
Table~\ref{tablethree}). Finally, we derive a simple expression for
the energy radiated in gravitational waves and find that this is
bounded between $\simeq 3.6\%$ and $\simeq 10\%$ for maximally
spinning binaries with spins anti-aligned or aligned with the orbital
angular momentum, respectively.

The plan of the paper is as follows: in Sect.~\ref{numerical
  simulations}, we recall very briefly the numerical set up and
illustrate the properties of the initial data used in the
simulations. Sect.~\ref{sec:GWO} is dedicated to the discussion of the
gravitational-wave observables used for the subsequent analysis, while
Sect.~\ref{sec:results} presents the results in terms of the SNR and
how this is influenced by higher-order modes. This Section also
contains a discussion of the match between the waveforms from
different binaries and an assessment of the accuracy of our results.
Sect.~\ref{sec:FF}, on the other hand, provides a brief discussion of
the analytic expressions we have found representing either the SNR or
the energy radiated in gravitational waves.  Finally, conclusions are
summarized in Sect.~\ref{sec:conclusions}.

\begin{table*}
\caption{\label{tableone}Binary sequences for which numerical
  simulations have been carried out, with various columns referring to
  the puncture initial location $\pm x/M$, the mass parameters
  $m_i/M$, the dimensionless spins $a_i$, and the normalized ADM mass
  ${\widetilde M_{_{\rm ADM}}}\equiv M_{_{\rm ADM}}/M$ measured at
  infinity.  Finally, the last four columns contain the numerical
  values of the energy radiated during the simulation using the two
  methods described in the text and the corresponding errors between
  them, as well as the error to the fitted values.}
\vspace{0.1cm}
\begin{ruledtabular}
\begin{tabular}{|l|ccccccc|rrcc|}
\hline
~					&
{$\pm x/M$} 				&
{$m_1/M$} 				&
{$m_2/M$} 				& 
{$a_1$} 				&
{$a_2$} 				&
{$(p_x,~p_y)_1=-(p_x,~p_y)_2$}      &
{${\widetilde M_{_{\rm ADM}}}$} 	&
{$E^{\rm NR}_{\rm rad}$}	 		&
{$E^{Q^{\times,+}}_{\rm rad}$}	&
{err.~($\%$)}                          &
{fit err.~($\%$)}                          \\ 
\hline
\hline
$r_0$   & $4.0000$ & $0.3997$ & $0.3998$ & $-0.600$     & $\pls 0.600 $ & $(0.002103,-0.112457)$ & $0.9880$ & $0.0366$ & $0.0356$ & $2.8$ & $1.6$ \\
$r_2$   & $4.0000$ & $0.3997$ & $0.4645$ & $-0.300$     & $\pls 0.600 $ & $(0.002024,-0.111106)$ & $0.9878$ & $0.0407$ & $0.0394$ & $3.3$ & $0.6$  \\
$r_4$   & $4.0000$ & $0.3998$ & $0.4825$ & $\pls 0.000$ & $\pls 0.600 $ & $(0.001958,~~0.001958)$ & $0.9876$ & $0.0459$ & $0.0445$ & $3.1$ & $1.9$\\
$r_6$   & $4.0000$ & $0.3999$ & $0.4645$ & $\pls 0.300$ & $\pls 0.600 $ & $(0.001901,-0.108648)$ & $0.9876$ & $0.0523$ & $0.0504$ & $3.8$ & $2.2$\\
\hline
$s_{-8}$& $5.0000$ & $0.3000$ & $0.3000$ & $-0.800$      & $-0.800$      & $(0.001300,-0.101736)$ & $0.9894$ & $0.0240$ & $0.0231$ & $3.8$ & $3.0$\\
$s_0$   & $4.0000$ & $0.4824$ & $0.4824$ & $\pls 0.000 $ & $\pls 0.000 $ & $(0.002088,-0.112349)$ & $0.9877$ & $0.0360$ & $0.0354$ & $1.7$ & $0.2$\\
$s_2$   & $4.0000$ & $0.4746$ & $0.4746$ & $\pls 0.200 $ & $\pls 0.200 $ & $(0.001994,-0.110624)$ & $0.9877$ & $0.0421$ & $0.0410$ & $2.7$ & $1.7$\\
$s_4$   & $4.0000$ & $0.4494$ & $0.4494$ & $\pls 0.400 $ & $\pls 0.400 $ & $(0.001917,-0.109022)$ & $0.9876$ & $0.0499$ & $0.0480$ & $4.0$ & $2.5$\\
$s_6$   & $4.0000$ & $0.4000$ & $0.4000$ & $\pls 0.600 $ & $\pls 0.600 $ & $(0.001860,-0.107537)$ & $0.9876$ & $0.0609$ & $0.0590$ & $3.2$ & $0.2$\\
$s_8$   & $4.0000$ & $0.4000$ & $0.4000$ & $\pls 0.800 $ & $\pls 0.800 $ & $(0.001816,-0.106162)$ & $0.9877$ & $0.0740$ & $0.0744$ & $0.5$ & $2.2$\\
\hline
$t_0$   & $4.0000$ & $0.3995$ & $0.3995$ & $-0.600$ & $-0.600$     & $(-0.002595,~0.118379)$ & $0.9886$ & $0.0249$ & $0.0243$ & $2.5$ & $1.1$\\
$t_1$   & $4.0000$ & $0.3996$ & $0.4641$ & $-0.600$ & $-0.300$     & $(-0.002431,~0.116748)$ & $0.9883$ & $0.0271$ & $0.0264$ & $2.7$ & $1.8$\\
$t_2$   & $4.0000$ & $0.3997$ & $0.4822$ & $-0.600$ & $\pls 0.000$ & $(-0.002298,~0.115219)$ & $0.9881$ & $0.0295$ & $0.0289$ & $2.1$ & $2.2$\\
$t_3$   & $4.0000$ & $0.3998$ & $0.4642$ & $-0.600$ & $\pls 0.300$ & $(-0.002189,~0.113790)$ & $0.9880$ & $0.0326$ & $0.0317$ & $2.8$ & $1.8$\\
\hline
$u_2$   & $4.0000$ & $0.4745$ & $0.4745$ & $-0.200$ & $\pls 0.200 $ & $(~0.002090,-0.112361)$ & $0.9878$ & $0.0361$ & $0.0354$ & $2.0$ & $0.2$\\
$u_4$   & $4.0000$ & $0.4492$ & $0.4494$ & $-0.400$ & $\pls 0.400 $ & $(~0.002095,-0.112398)$ & $0.9879$ & $0.0363$ & $0.0355$ & $2.3$ & $0.7$\\
$u_8$   & $4.0000$ & $0.2999$ & $0.2999$ & $-0.800$ & $\pls 0.800 $ & $(~0.002114,-0.112539)$ & $0.9883$ & $0.0374$ & $0.0363$ & $3.0$ & $3.7$\\
\hline
\end{tabular} \\
\end{ruledtabular}
\end{table*}

\section{Numerical Setup and Initial Data}
\label{numerical simulations}

The numerical simulations have been carried out using the
\texttt{CCATIE} code, a three-dimensional finite-differencing code
solving a conformal-traceless ``$3+1$'' BSSNOK formulation of the Einstein
equations~\cite{Pollney:2007ss} using the \texttt{Cactus}
Computational Toolkit~\citep{cactusweb} and the
\texttt{Carpet}~\citep{Schnetter-etal-03b} adaptive mesh-refinement
driver. The main features of the code have been presented in several
papers, and recently reviewed in~\citet{Pollney:2007ss}. The code
implements the ``moving-punctures'' technique to represent dynamical
black holes following ~\citep{Baker:2006yw,Campanelli:2005dd} (see
also \cite{Hannam:2006vv,Thornburg-etal-2007a}), which has proven to
be a robust way to evolve black-hole spacetimes.

For compactness we will not report here the details of the formulation
of the Einstein equations solved or the form of the gauge conditions
adopted. All of these aspects are discussed in great detail
in~\cite{Pollney:2007ss}, to which we refer the interested
reader. More specific to these simulations, however, is the numerical
grid setup.  In the results presented below we have used 9 levels of
mesh refinement with a fine-grid resolution of $\Delta x/M=0.02$ and
fourth-order finite differencing.  The
wave-zone grid has a resolution of $\Delta x/M=0.128$ and extends from
$r=24\,M$ to $r=180\,M$, in which our wave extraction is carried
out. The outer (coarsest) grid extends to a spatial position which is
$819.2\,M$ in each coordinate direction. Furthermore, because the
black holes spins are all directed along the $z$-axis of our Cartesian
grids, it is possible to use a reflection symmetry condition across
the $z=0$ plane.

\begin{figure}
 \begin{center}
   \scalebox{0.425}{\includegraphics{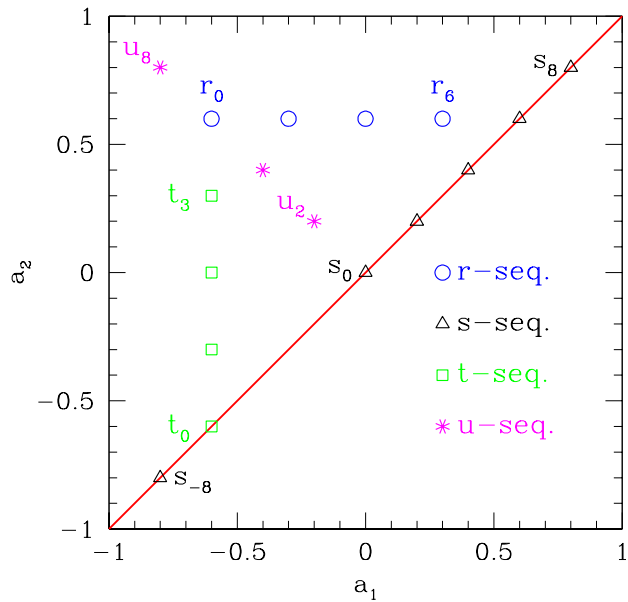}}
 \end{center}
 \caption{Schematic representation in the $(a_1,\,a_2)$ plane, also
   referred to as the ``spin diagram'', of the initial data collected
   in Table~\ref{tableone}. These sequences cover most important
   portions of the space of parameters which is symmetric with respect
   to the $a_1=a_2$ diagonal.}
  \label{fig:seq}
\end{figure}

The initial data are constructed applying the ``puncture'' method
\cite{Brandt97b,Bowen80,Beig:1993gt,Dain:2001ry} as described
in~\cite{Ansorg:2004ds}. We have considered four different sequences
labelled as \textit{``r'', ``s,'' ``t''}, and \textit{``u''} along
straight lines in the $(a_1, a_2)$ parameter space, also referred to
as the ``spin diagram'' (\cf Table~\ref{tableone} for details). As
shown in Fig.~\ref{fig:seq}, these sequences allow us to cover the
most important portions of the space of parameters which, we recall,
is symmetric with respect to the $a_1=a_2$ diagonal.

We note that similar sequences have also been considered
in~\cite{Koppitz-etal-2007aa,Pollney:2007ss, Rezzolla-etal-2007,
  Rezzolla-etal-2007b, Rezzolla-etal-2007c} but have here been
recalculated both using a higher resolution and with improved initial
orbital parameters. More specifically, we use post-Newtonian (PN)
evolutions following the scheme outlined in~\cite{Husa:2007rh}, which
provides a straightforward prescription for initial-data parameters
with small initial eccentricity, and which can be interpreted as part
of the process of matching our numerical calculations to the inspiral
described by the PN approximations.  The free parameters to be chosen
for the puncture initial data are therefore: the puncture coordinate
locations ${\boldsymbol C}_i$, the puncture bare mass parameters $m_i$, 
the linear momenta ${\boldsymbol p}_i$, and the individual spins
${\boldsymbol S}_i$.  The initial parameters for all of the binaries
considered are collected in the left part of Table~\ref{tableone}. The
initial separations are fixed at $D=8\,M$, where $M$ is the total
initial black hole mass, chosen as $M=1$ (note that the initial ADM
mass of the spacetime is not exactly 1 due to the binding energy of
the black holes), while the individual asymptotic initial black hole
masses are therefore $M_i = 1/2$.  The only exception is for the
binary $s_{-8}$, for which $D=10\,M$.

\section{Gravitational-Wave Observables}
\label{sec:GWO}

In this Section we discuss the gravitational-wave observables that
have been studied from the sample reported in Table~\ref{tableone} and
how these have been used to compute the radiated energy, the SNR, the
horizon distances and the event rates.

\subsection{Numerical-Relativity (NR) waveforms}

Although the \texttt{CCATIE} code computes the gravitational waveforms
either via the Newman-Penrose curvature scalar $\Psi_4$ or via
gauge-invariant metric perturbations on a Schwarzschild background,
the analysis carried hereafter will be made in terms of the
latter. While the two prescriptions yield, in fact, estimates which
are in very good agreement with each other and with differences below
$2\%$ (see discussion in~\cite{Pollney:2007ss}), we have found that
the results obtained using gauge-invariant quantities have a smaller
numerical error, and are thus preferable.

More specifically, we compute the gravitational-wave amplitudes
$h^{+}_{\ell m}$ and $h^{\times}_{\ell m}$ in terms of the even and
odd master functions $Q^{+}_{\ell m}$ and $Q^{\times}_{\ell m}$
via the relations~\cite{Nagar05}
\begin{equation} 
h_{\ell m}(t) = 
h^{+}_{\ell m}(t) - {\rm i} h^{\times}_{\ell m}(t) =
Q^{+}_{\ell m}(t) - {\rm i} \int_{-\infty}^{t} dt' Q^{\times}_{\ell
  m}(t')\,,  
\end{equation}
where the gauge-invariant perturbations are typically extracted at a
radius of $r_{_{\rm E}}=160M$ (see Sec.~\ref{sec:amplitude_accuracy}
for a discussion of the accuracy of our measurements and
ref.~\cite{Pollney:2007ss} for a comparison among different extraction
radii).  

As mentioned before, all our binaries [but $s_{-8}$] have initial
separations of $D=8.0M$ [$D=10.0M$], which, in the parameter space
that we have considered, leads to a maximum initial frequency of the
numerical waveforms, that is $\omega_{\rm ini}=0.084/M$. Depending
therefore on the mass $M$, such an initial frequency can be greater
than the lower cut-off frequency of the detector $\omega_{\rm co}$ for
a given source at an arbitrary distance. Because we expect that for
most masses $\omega_{\rm co}$ will be smaller than $\omega_{\rm ini}$,
we need to provide additional information about the gravitational-wave
signal in the frequency range between $\omega_{\rm co}$ and
$\omega_{\rm ini}$. This can be accomplished by ``gluing'' the NR
waveform with a PN part as discussed in the next Section.

The values of the initial frequencies and of the associated minimum
masses $M_{\rm min}$ for each of the detectors considered are reported
in Table~\ref{tabletwo}.

\begin{table}
\caption{\label{tabletwo}Initial instantaneous frequencies
  $M\omega_{\rm ini}$ and associated minimum masses $M_{\rm min}$ of
  the NR waveforms for the different models and for each detector
  according to the corresponding lower cut-off frequency (\ie at $30$
  Hz for Virgo, at $40$ Hz for eLIGO, at $10$ Hz for AdLIGO/AdVirgo,
  and at $10^{-4}$ Hz for LISA). All the values for the masses are in
  units of solar masses.}
\vspace{0.1cm}
\begin{ruledtabular}
\begin{tabular}{|l|c|cccc|}
\hline
~			        &
$M\omega_{\rm ini}$               &
$M_{\rm min}$                     &
$M_{\rm min}$                     &
$M_{\rm min}$                     &
$M_{\rm min}$                     \\
~			        &
~			        &
Virgo                           &
eLIGO                           &
AdLIGO/AdVirgo                  &
LISA                           \\
\hline
\hline
$r_0$   & $0.080$ & $86.2$ & $64.6$ & $258.5$ & $2.58\times 10^7$ \\
$r_2$   & $0.078$ & $84.0$ & $63.0$ & $252.0$ & $2.52\times 10^7$ \\ 
$r_4$   & $0.077$ & $82.9$ & $62.2$ & $248.8$ & $2.49\times 10^7$ \\
$r_6$   & $0.076$ & $81.8$ & $61.4$ & $245.5$ & $2.46\times 10^7$ \\  
\hline                     
$s_{-8}$ & $0.060$ & $64.6$ & $48.4$ & $193.8$ & $1.93\times 10^7$\\
$s_0$   & $0.080$ & $86.2$ & $64.6$ & $258.5$ & $2.58\times 10^7$\\
$s_2$   & $0.078$ & $84.0$ & $63.0$ & $252.0$ & $2.52\times 10^7$\\  
$s_4$   & $0.076$ & $81.8$ & $61.4$ & $245.5$ & $2.46\times 10^7$\\  
$s_6$   & $0.075$ & $80.8$ & $60.6$ & $242.3$ & $2.42\times 10^7$\\  
$s_8$   & $0.073$ & $78.6$ & $59.0$ & $235.8$ & $2.36\times 10^7$\\  
\hline                     
$t_0$   & $0.084$ & $90.5$ & $67.8$ & $271.4$ & $2.71\times 10^7$\\  
$t_1$   & $0.083$ & $89.4$ & $67.0$ & $268.2$ & $2.68\times 10^7$\\  
$t_2$   & $0.082$ & $88.3$ & $66.2$ & $264.9$ & $2.65\times 10^7$\\  
$t_3$   & $0.081$ & $87.2$ & $65.4$ & $261.7$ & $2.62\times 10^7$\\  
\hline                     
$u_2$   & $0.080$ & $86.2$ & $64.6$ & $258.5$ & $2.58\times 10^7$\\  
$u_4$   & $0.080$ & $86.2$ & $64.6$ & $258.5$ & $2.58\times 10^7$\\  
$u_8$   & $0.080$ & $86.2$ & $64.6$ & $258.5$ & $2.58\times 10^7$\\  
\hline
\end{tabular} \\
\end{ruledtabular}
\end{table}

\subsection{Matching PN and NR waveform amplitudes}

The existence of a cut-off mass set by the initial frequency of the NR
simulations would clearly restrict the validity of our considerations
to large masses only. To counter this and thus include also binaries
with smaller masses, we account for the early inspiral phase by
describing it via PN approximations. To produce the PN waveforms, and
the PN energy that we are using directly in Sec. \ref{sec:fitErad}, 
we have used
the spinning TaylorT1 approximant used
in Hannam et al.~\cite{Hannam:2007wf}, and which is based on the PN expressions
described in~\cite{Damour:2001bu, Blanchet:2001ax, Blanchet:2004ek,
  Kidder:1995zr, Poisson:1997ha, Alvi:2001mx, Blanchet:2006gy,
  Faye:2006gx}. The choice of TaylorT1 is motivated by
that fact, that in~\cite{Hannam:2007wf} it is found to be more
robust in the spinning case than the TaylorT4 approximant, which
was previously found to yield excellent
results in the nonspinning case~\cite{Boyle:2007ft}
(see \eg \cite{Boyle:2007ft} for a
comparison of different techniques to obtain the gravitational-wave
phase information for quasi-circular inspiral).
These waveforms are 3.5 PN accurate in the nonspinning phase, 
and 2.5 PN accurate in the spin-dependent terms entering the
phasing.
The gravitational-wave
amplitudes, on the other hand, have been computed according to
ref.~\cite{Blanchet:2008je} (see also \cite{Kidder:2007rt}) to the
highest PN order that is currently known for each of the spherical
harmonic modes that we use.

\begin{figure}
 \begin{center}
   \scalebox{0.425}{\includegraphics{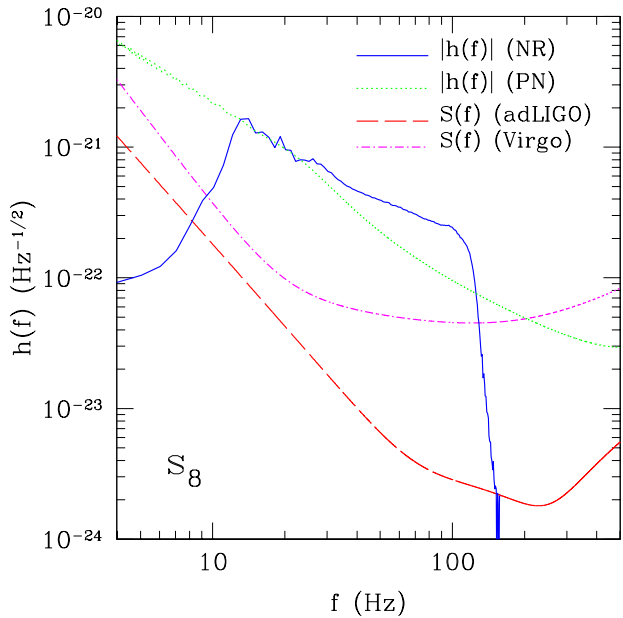}}
 \end{center}
 \caption{Noise strain for the Advanced LIGO and Virgo detectors and
   the Fourier-transformed amplitude of the PN and NR waveform at
   $\theta=0, \phi=0$ for a total mass $M=200\,M_{\odot}$ at a
   distance $d=100\,\mpc$ for the maximally spinning model $s_8$. The
   glueing frequency is at $f_{\rm glue}=27.14$ Hz. }
   \label{fig:SNRadvLIGO}
\end{figure}

A phase-coherent construction of hybrid PN-NR waveforms is rather
delicate, and has not yet been achieved for the higher spherical
harmonic modes we use here (see~\cite{Ajith:2007qp,Ajith:2007kx} for
some recent work in the case of nonspinning binaries). However, for
the present purpose of computing the SNR and the radiated energies,
such a construction in the time domain is not necessary and all of the
relevant work can be done much more simply in the frequency domain. In
practice, we Fourier transform the PN and NR waveforms and ``glue''
them together at a suitable ``glueing'' frequency $\omega_{\rm
  glue}$. Since the SNR depends only on the amplitude of the waveform,
[\cf eq.~\eqref{SignalToNoiseRatio}], it is not necessary to match the
PN-waveform in the phase.  This greatly simplifies the process of
waveform matching and basically reduces to a simple check of the
amplitude matching to address the error of the mismatch.  Indeed, we
have found that without any parameter adjustment, the PN-waveform
amplitudes match rather well with the inspiral part of the
NR-waveforms, and result in an error which is usually $\approx 1.5\%$
and in the worst case $\approx 4.0\%$ for the binary configuration
$t_0$.  The only care which is important to pay in the time-domain
analysis, and in order to limit the noise artifacts in the
Fourier-transformed amplitudes, is the use of a windowing function
(\eg a hyperbolic tangent) to smoothly blend the waveform to zero
before the initial burst of spurious radiation and after the ringdown,
in order to limit spurious oscillations in the Fourier-transformed
waveform. A representative example is shown in
Fig.~\ref{fig:SNRadvLIGO}, where we report the noise strain for the
Virgo and Advanced LIGO detectors, together with the
Fourier-transformed amplitude of the PN and NR waveform for the
maximally spinning model $s_8$. The waveform is assumed to be observed
at $\theta=0, \phi=0$ for a total mass $M=200\,M_{\odot}$ and from a
distance $d=100\,\mpc$. The glueing frequency in this case is at
$f_{\rm glue}=\omega_{\rm glue}/(2\pi)=27.14$ Hz.

Since each $\ell, m$ mode of the gravitational-wave
field will have a different initial frequency, we need to make
sure that they are all properly taken into account when determining
the glueing frequency, so that, at least in principle
\begin{equation}
\omega_{\rm glue}\geq\max_{\ell, m} (\omega_{\rm ini})_{\ell m}\,.
\end{equation}
In practice, the initial frequency of our highest mode, $\ell=4, m=4$,
has an initial frequency $(\omega_{\rm ini})_{4 4}=2(\omega_{\rm
  ini})_{2 2}$ \, .  As a result, we select the glueing frequency
according to the binary configuration with the largest initial
frequency, \ie the binary $t_0$, and take $\omega_{\rm glue}=
2(\omega_{\rm ini})_{2 2} = 0.168/M$. We also measure how sensitive
this choice is, by considering how the results are affected when
choosing instead $\omega_{\rm glue} \pm \Delta\omega$, with
$\Delta\omega \ll \omega_{\rm glue}$. More specifically, for $\Delta
\omega = 0.01/M$ we find a maximal difference in the computed SNR of
$\sim 2.0\%$ over all configurations and all masses. Note that such a
difference affects equally the maximum and averaged SNRs (see
Sect.~\ref{sec:SNR} for a discussion on these two different measures
of the SNR). Furthermore, a change of $\Delta \omega$ in $\omega_{\rm
  glue}$ affects only marginally the relative difference between SNRs
computed by including modes up to $\ell=2$ and $\ell=4$, and also in
this case the differences are $\sim 2.0\%$. Overall, therefore, the
uncertainties introduced by the choice of $\omega_{\rm glue}$ are much
smaller than the typical error at which we report the SNRs.

\subsection{Radiated Energy}\label{subsec:RadiatedEnergy}

Since the total energy must be conserved, we can use the
radiated energy as an important tool to verify the accuracy of the
gravitational-wave amplitude and thus the overall precision of our
calculations. More specifically, because it is straightforward to
determine the initial and the final total mass, it is also
straightforward to compare the difference in the two with the radiated
energy. In practice, we compute the initial mass of the system as
$M_{{\rm ini}} = {\widetilde M}_{_{\rm ADM}}$, while the final mass of
the merger remnant $M_{{\rm fin}}$ is deduced from the properties of
the apparent horizon within the isolated-horizon formalism as first
discussed in~\cite{Dreyer-etal-2002-isolated-horizons} and then
extensively investigated in~\cite{Baiotti04}. The radiated energy is
then simply given by the difference
\begin{equation} 
\label{first_Erad_method} 
E^{\rm NR}_{{\rm rad}}=M_{_{\rm ADM}}-M_{{\rm fin}}\,, 
\end{equation} 
and should be equal to the energy that has been radiated through
gravitational waves during the simulation~\cite{Nagar05}
\begin{equation}
E^{Q^{\times,+}}_{{\rm rad}}
= \frac{1}{32\pi} \sum_{\ell, m} 
\int_{0}^{t}dt' \left( \left| \frac{dQ^+_{\ell m}}{dt} \right|^2 + 
\left| Q^{\times}_{\ell m}\right|^2  \right)\,. 
\label{Erad_sim}
\end{equation}
Overall, we have found that for all binaries the difference between
$E_{{\rm rad}}$ and $E^{Q^{\times,+}}_{{\rm rad}}$ is between $\sim
0.5\%$ and $\sim 4.0\%$ and a detailed comparison of the numerical
values is reported in Table~\ref{tableone}.  In
Sect.~\ref{sec:fitErad} we will discuss an analytic fit to the
computed data that provides a simple-to-use measure of the amount of
mass radiated during the inspiral, merger and ringdown as a function
of the initial spins.

\subsection{SNR, Horizon Distances and Event Rates}
\label{sec:SNR}

Following ref.~\cite{Flanagan:1998a}, we define the SNR, $\rho$, for
matched-filtering searches as
\begin{equation}
\rho^2 \equiv \left(\frac{S}{N}\right)^2_{\rm matched} = 
4 \int_0^\infty\frac{|\tilde{h}(f)|^2}{S_h(f)} df \,,
\label{SignalToNoiseRatio}
\end{equation}
where $\tilde{h}(f)$ is the Fourier transform of the time domain
gravitational-wave signal $h(t)$, defined in the continuum as
\begin{equation}
\tilde{h}(f) = \int_{-\infty}^{\infty} h(t) e^{- 2 \pi {\rm i} f t} dt\,,
\end{equation}
and $S_h(f)$ is the noise power spectral density for a given
detector. Hereafter we will consider the $S_h(f)$ for the ground-based
detectors LIGO, enhanced LIGO, advanced LIGO and Virgo, as well as the
space-bound LISA interferometer. [The associated noise power spectral
  densities are reported in Appendix~\ref{sec:appendix_a}.]

Note that since the SNR (\ref{SignalToNoiseRatio}) depends on the
angle from the source to the detector, it is useful to introduce the
angle-averaged SNR $\langle\rho^2\rangle$, which can be computed
straightforwardly after decomposing the gravitational-wave signal in
terms of spherical harmonic modes. More specifically, using the
orthonormality of the spin-weighted spherical harmonic basis
$_sY_{\ell m}$, the \textit{``angle-averaged''} SNR
\begin{equation}
\rho_{\rm avg} \equiv \langle\rho^2\rangle \equiv 
\frac{1}{\pi} \int d\Omega \int df \frac{\left|\sum_{\ell
    m}\tilde{h}_{\ell m}(f)\, {}_{-2}Y_{\ell
    m}(\Omega)\right|^2}{S_h(f)}
\end{equation}
can be written as a simple sum of integrals of the absolute squares of
the Fourier-transformed modes $\tilde{h}_{\ell m}(f)$
\begin{equation}
\rho_{\rm avg} = \frac{1}{\pi}
\sum_{\ell m} \int df \frac{|\tilde{h}_{\ell m}(f)|^2}{S_h(f)}\,,
\end{equation}
and hence it can be evaluated straightforwardly. For each binary,
distance and mass, we have calculated both the \textit{``maximum''}
SNR $\rho_{\rm max}$ for an optimally oriented detector, \ie the SNR
for a detector oriented such that it measures only the $+$
polarization of the gravitational-wave signal, and the averaged SNR.
Here the mass is always meant to be the \textit{redshifted} total
mass, \ie $(1+z)M_{\rm source}$, where $z$ is the redshift and $M_{\rm
  source}$ is the mass at the source. For sources at small distances,
\ie less than $100\,\mpc$, then $z \lesssim 0.024$ and hence $M \simeq
M_{\rm source}$ to within a few percent. Identical results would have
been obtained if we had considered the $\times$ polarization.

It is worth noting that if the gravitational-wave signal is modeled
simply through the dominant $\ell=2=m$ mode (or in our case via a
superposition $\ell=2=\pm m$)~\footnote{Note that in our binary
  configurations due to symmetry, we always have $h_{\ell m}=h_{\ell
    -m}$}, the maximum SNR can be deduced from the average SNR after
exploiting the properties of the spin-weighted spherical harmonic
$_{-2}Y_{22}$ and $_{-2}Y_{2-2}$, namely,
\begin{eqnarray}
\label{eq:SNRR}
\rho_{\rm max} &=& \sqrt{5\rho^2_{\rm avg}(\ell=2,m=2)}   \\
             &=& \sqrt{\frac{5}{2}\rho^2_{\rm avg}(\ell=2,m=\pm2)}\,.
\end{eqnarray}
However, such a relation is no longer true when including modes with
$\ell>2,$ and the relation between the maximum and the averaged value
of the SNR can only be determined numerically.

When computing the SNR, a reference distance needs to be fixed and we
have set such a distance to be $d_\rho=100\, {\rm Mpc}$. The results
of the SNR at $d_{\rho}$ across the spin diagram can then be recast in
terms of an \textit{``horizon distance''}, namely the distance at
which a given binary system with redshifted mass $M$ has an SNR equal
to a threshold for detectability, which we chose to be $\rho=8$, as
customary for ground-based detectors. The horizon distance is then
simply defined as
\begin{equation}
\label{eq:dH}
d_H = d_\rho \, \left(\frac{\rho(d=d_\rho)}{8}\right)\, {\rm Mpc} \,.
\end{equation}
The quantity $d_H$ is clearly equivalent to the SNR but has the
advantage to provide, at least for detectors not operating at large
SNRs, a simple estimate of the increase in the relative event rate $R$
as
\begin{equation}
\label{eq:R}
R \sim  \left(\frac{d_H}{d_{H, a=-1}}\right)^3\,,
\end{equation}
where $d_{H, a=-1}$ is the horizon distance of the configuration with
lowest SNR, \ie which belongs to the extrapolated case $a=-1$.
Although simple, this formula requires a
caveat. Expression~\eqref{eq:R} is valid as an equality only for small
horizon distances, namely those for which the redshift is
negligible. This is because at large redshifts the observed masses
would differ considerably from the masses at the source. In other
words, at large redshifts the horizon distances would be different not
only because of the spin, but also because the masses at the sources
would be intrinsically different. This clearly impacts the deduced
event rate as defined in~\eqref{eq:R}, which considers only the
contributions coming from the spin. Hence, for large redshifts the
event rate $R$ defined here serves only as a lower limit for masses
larger than the optimal one and, vice versa, as an upper bound for masses
smaller than the optimal.

\begin{figure*}
 \begin{center}
   \scalebox{0.35}{\includegraphics[angle=-90]{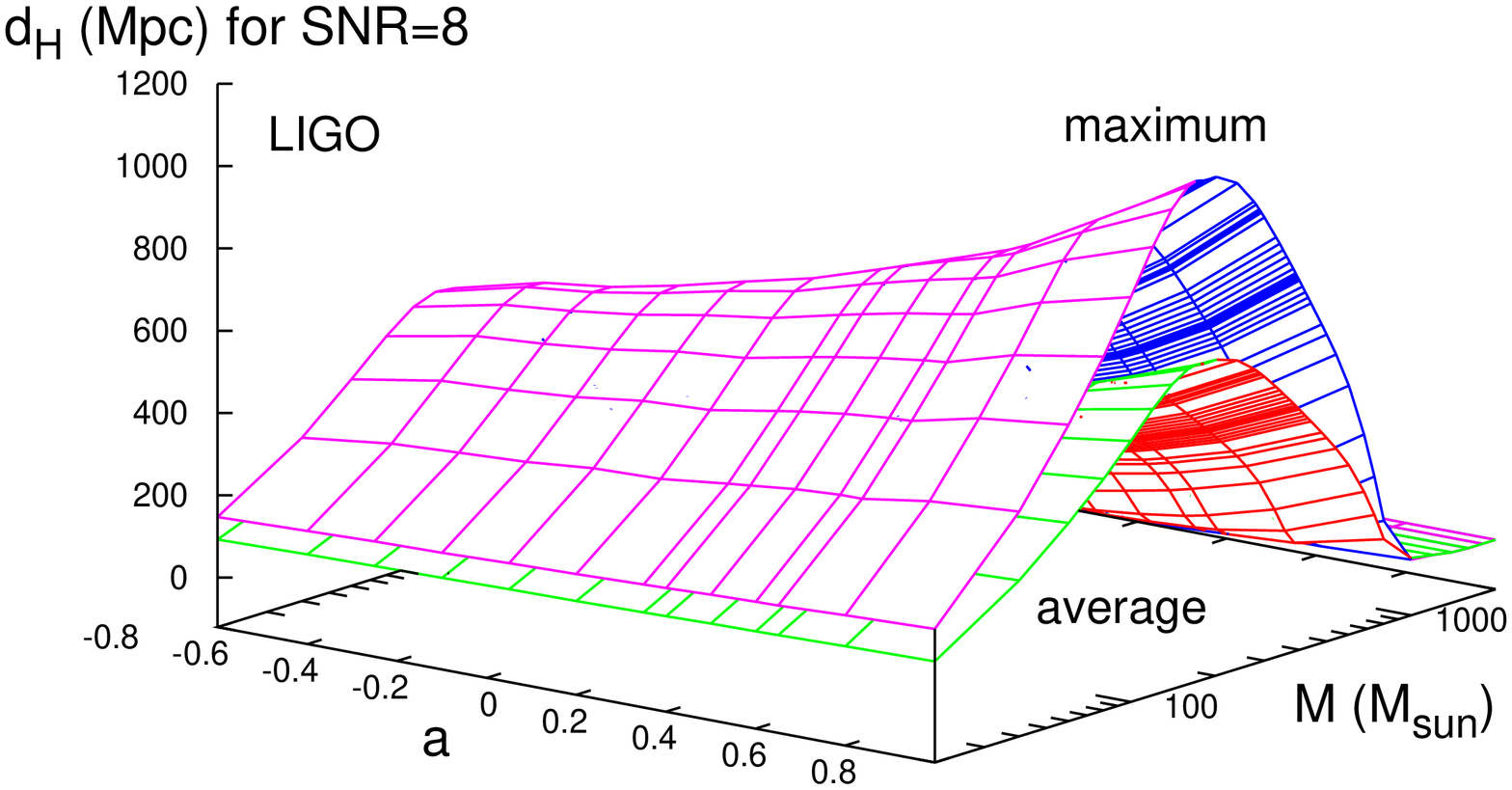}}
   \scalebox{0.35}{\includegraphics[angle=-90]{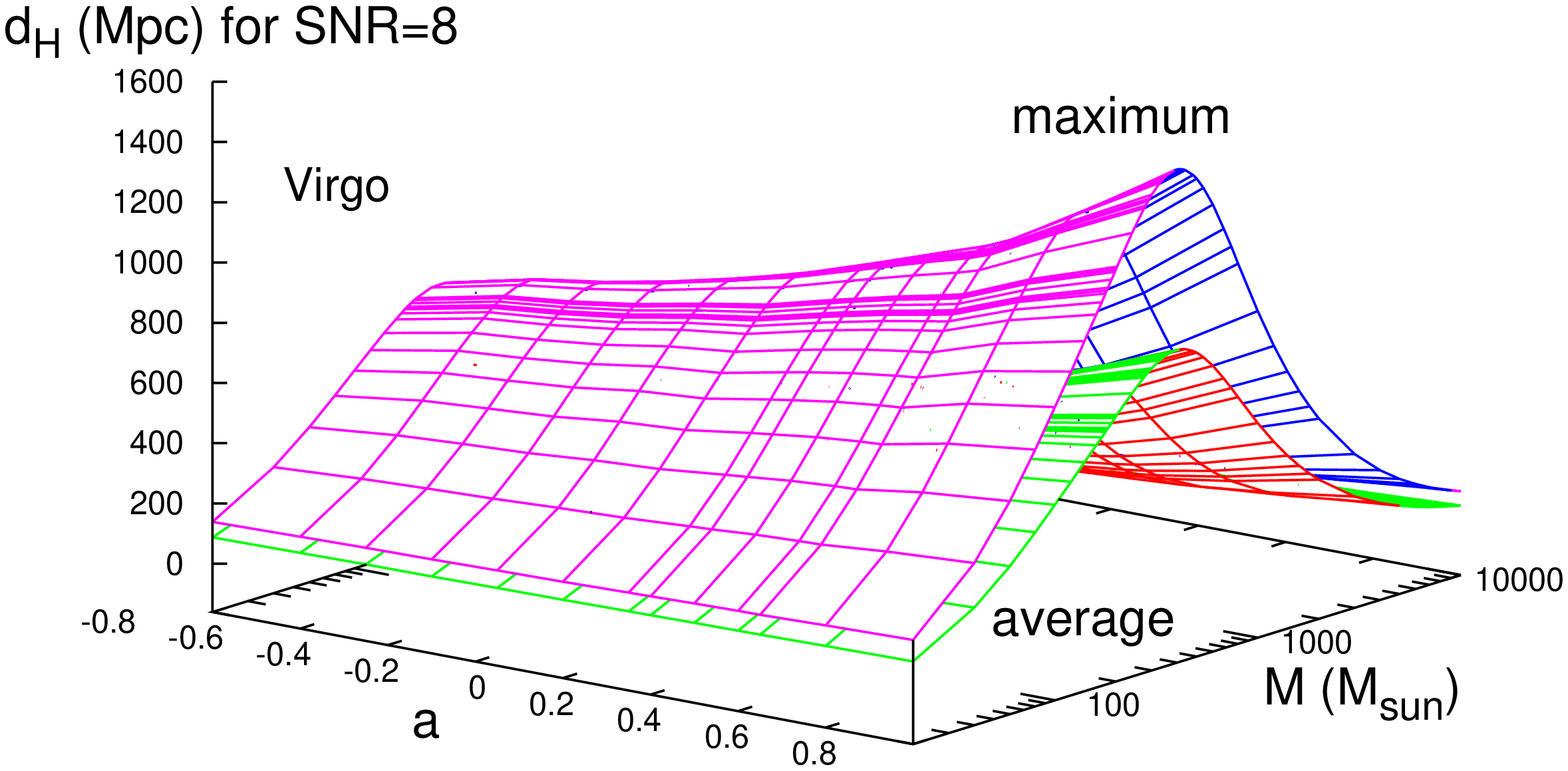}}
   \scalebox{0.35}{\includegraphics[angle=-90]{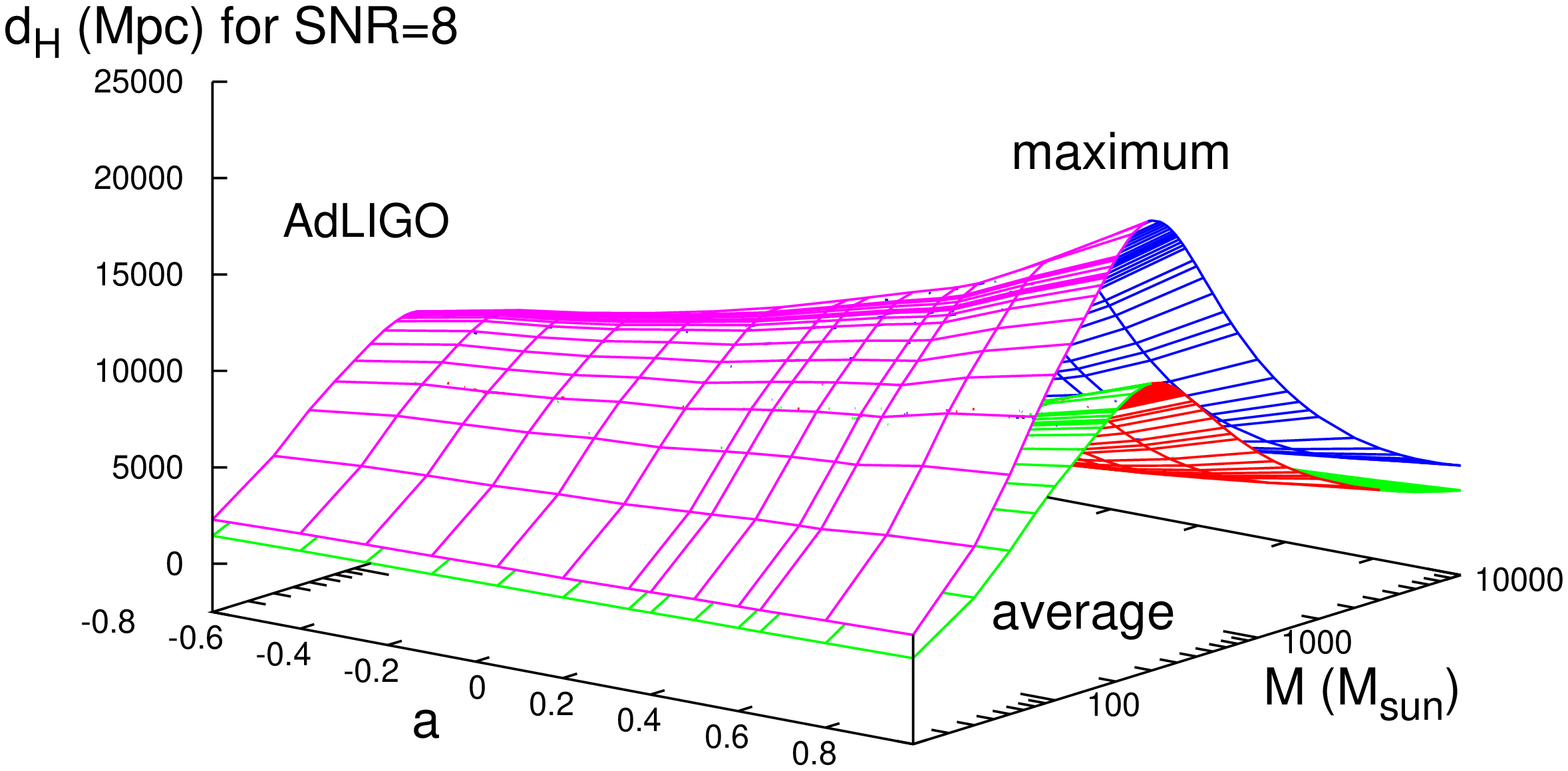}}
   \scalebox{0.35}{\includegraphics[angle=-90]{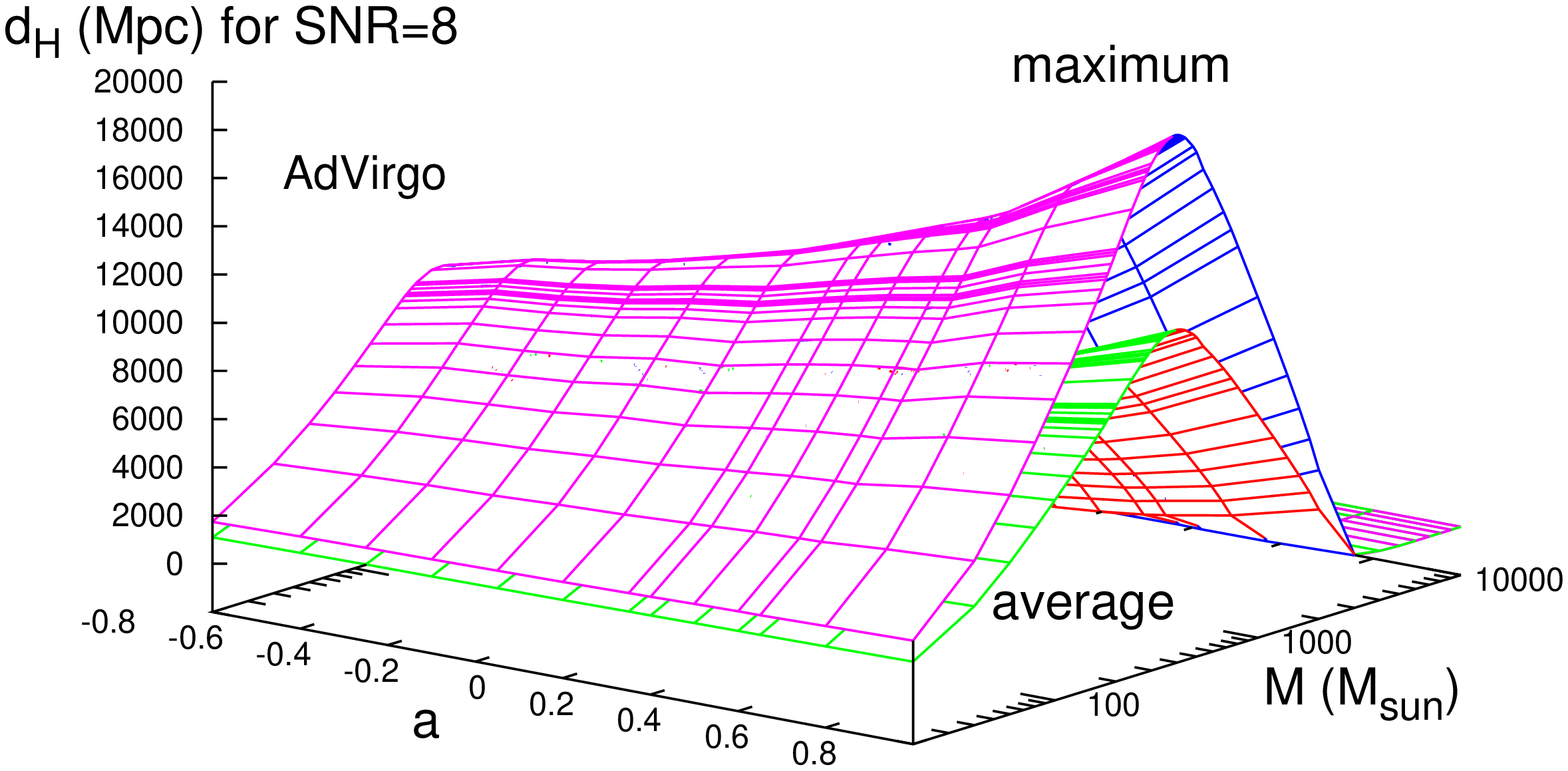}}
 \end{center}
 \caption{Averaged and maximum horizon distance $d_H=d_H(a, M)$ for
   the LIGO detector (top left panel), for the Virgo detector (top
   right panel), and for the advanced versions of both detectors
   (bottom left and right panels, respectively). The horizon distance
   has been computed at a reference SNR $\rho=8.0$.}
   \label{fig:SNRLIGO}
\end{figure*}

To fix the ideas, let us consider a concrete example. Let us assume
that we have calculated the horizon distance for a binary with $a=-1$
which, as can be deduced from Fig.~\ref{fig:max-avg-SNR} and will be
discussed in the next Section, will lead to the smallest SNR for a
given detector. We also assume that this binary has a mass at the
detector which is smaller than the optimal one. Let us now consider a
binary with the same mass at the detector but with $a>-1$; this binary
will clearly lead to a larger SNR but because the masses at the
detector are the same, the mass of the binary with $a>-1$ will be
(because of the redshift) smaller at the source. As a result, its
horizon distance will be overestimated, and hence the event rate
coming from~\eqref{eq:R} only an upper bound. A similar argument for
masses larger than the optimal one would instead lead to the
conclusion that the event rate $R$ is only a lower bound.

\section{Results}
\label{sec:results}

In what follows we discuss the results obtained in terms of the SNR
and how this is influenced by higher-order modes. We also discuss the
match between the waveforms from different binaries and an assessment
of the accuracy of our results.

\subsection{Horizon distances and SNRs}

The results of the analysis discussed above are nicely summarized in
Fig.~\ref{fig:SNRLIGO}, which shows the averaged and maximum horizon
distance $d_H=d_H(a, M)$ for some of the detectors considered. As
mentioned above, the horizon distance has been computed at a reference
SNR $\rho=8.0,$ and is parametrized in terms of the total mass of the
system (in solar masses) and of the average dimensionless spin ``$a$''
as projected along the orbital angular momentum $\boldsymbol L$
\begin{equation}
a \equiv \frac{1}{2}(\boldsymbol{a}_1  + \boldsymbol{a}_2 )\cdot 
	\boldsymbol{\hat{L}}
        = \frac{1}{2}(\boldsymbol{a}_1  + \boldsymbol{a}_2 )\cdot 
	\boldsymbol{e}_z\,, 
\end{equation}
where $\boldsymbol{\hat{L}} \equiv \boldsymbol{L}/|\boldsymbol{L}|$,
and the orbital plane has been chosen to coincide with the $(x,\,y)$
plane of our Cartesian coordinate system. More specifically, the top
left panel of Fig.~\ref{fig:SNRLIGO} refers to the LIGO detector, the
top right panel to the Virgo detector, while the lower left and right
panels refer to the advanced versions of both detectors, respectively.

\begin{figure}
 \begin{center}
   \scalebox{0.425}{\includegraphics{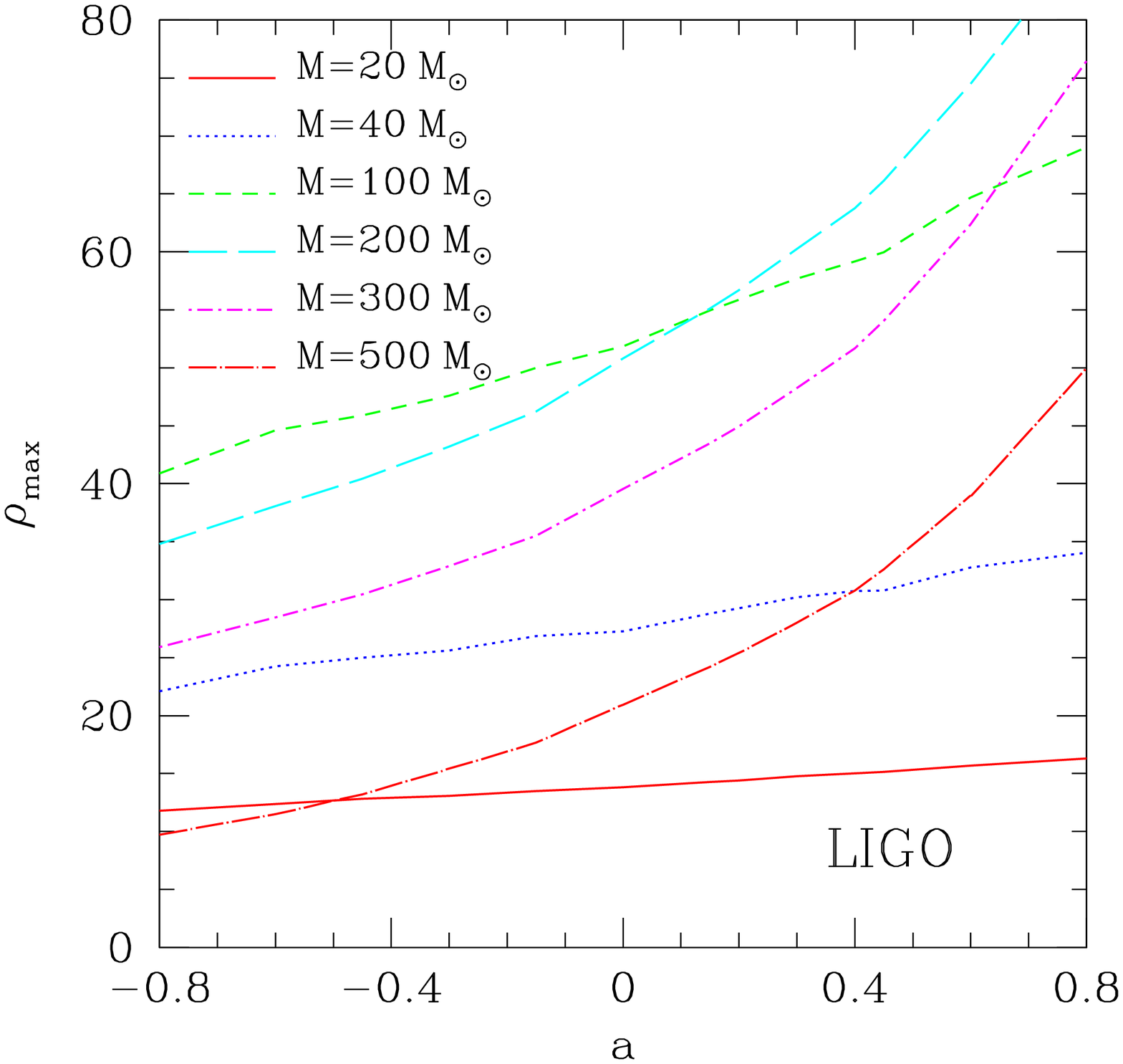}} 
 \end{center}
 \caption{Maximum SNR $\rho_{\rm max}=\rho(a,\,M)$ for the LIGO
   detector for a given set of masses at a distance
   $d=100\,\mpc$. Note that the growth of $\rho_{\rm max}$ with $a$ is
   very well described with a low-order polynomial which is of $4$th
   order for the optimal mass (\cf~discussion in
   Sect.~\ref{sec:fitSNR}). Note also that the dependence on $a$
   becomes stronger for masses $M>200 \,M_{\odot}$, for which the
   NR-part of the waveform and hence the plunge and ringdown phase
   dominate. In these cases, the SNR is more then doubled between
   $a=-1$ and $a=+1$.}
   \label{fig:max-avg-SNR}
\end{figure}

\begin{figure}
 \begin{center}
   \scalebox{0.375}{\includegraphics[angle=-90]{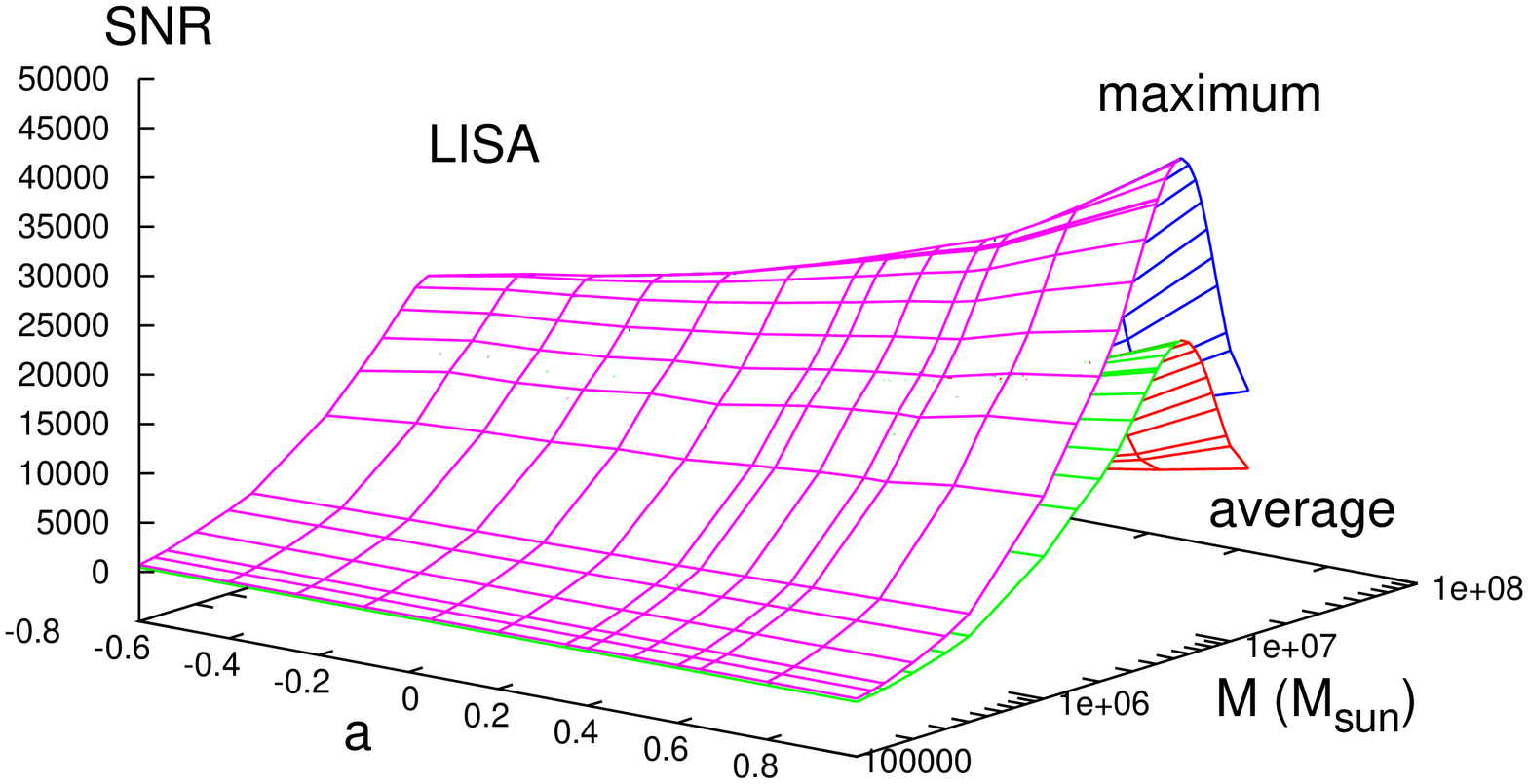}}
 \end{center}
 \caption{Averaged and maximum SNR $\rho=\rho(a, M)$ for the planned
   LISA mission and for sources at $d=6.4\,\mathrm{Gpc}\ (z=1)$.}
   \label{fig:SNRLISA}
\end{figure}

While quite self-explanatory, these panels deserve some
comments. First, as expected, the maximum SNR is always larger than
the average one but the difference between the two is not constant,
changing both with the total dimensionless spin $a$ and with the total
mass $M$. Second, for any fixed value of $a,$ the horizon distance
(and hence the SNR) grows steeply to a maximum mass and then rapidly
decreases to very small values of $\sim {\cal O}(1)$. Clearly, this
reflects the existence of a sweet-spot in the sensitivity curve of all
detectors.  Third, for any value of $a$, the maximum horizon
distance/SNR also marks the ``optimal mass'' for the binary $M_{\rm
  opt}$, namely the mass of the binary whose inspiral and merger is
optimally tuned with the given detector and hence can be seen from
further away. Note that the differences between the maximum and
average SNR are largest in the neighborhood of the optimal
mass. Fourth, the configuration with spins parallel and aligned to the
orbital angular momentum are generically ``louder'' than those with
spins parallel but antialigned with the orbital angular momentum, with
the binaries having $a=\pm 1$ being the ``loudest'' and ``quietest'',
respectively; this is essentially the answer to question \textit{(i)}
in the Introduction.~\footnote{This behaviour can be easily understood
  in terms of the orbital dynamics: the binaries with larger total
  angular momentum will have a larger number of cycles and hence a
  larger SNR}. Fifth, in the cases of the LIGO and advanced Virgo
detectors the horizon distance is essentially zero at cut-off masses
which are $\sim 900\,M_{\odot}$ and $\sim 3000\,M_{\odot}$,
respectively. Sixth, for any fixed value of the total mass, the SNR
grows with $a$ and, as we will discuss later on, this growth is very
well described with a polynomial of $4$th order (\cf~discussion in
Sect.~\ref{sec:fitSNR}). This is shown more clearly in
Fig.~\ref{fig:max-avg-SNR}, which reports the maximum SNR $\rho_{\rm
  max}$ for the LIGO detector and for a given set of masses at a
distance $d=100\,\mpc$. Note that the growth of $\rho_{\rm max}$ with
$a$ becomes steeper for masses $M>200 \,M_{\odot}$, for which the
NR-part of the waveform and hence the plunge and ringdown phase
dominates. In these cases, the SNR is more then doubled between $a=-1$
and $a=+1$. Finally, when going from the present LIGO/Virgo detectors
to their advanced versions, the average horizon distances go from
$\sim 600/800\,\mpc$ to $\sim 10^4/1.2\times10^4\,\mpc$, thus with an
observational \textit{volume} of the Universe that is increased by a
factor of $\sim 5000/3000$, respectively. Note that if we assume a
Hubble radius of $\sim 4.1\,{\rm Gpc}$, both detectors would
effectively detect binaries within a large range of masses (\eg $60
\lesssim M/M_{\odot} \lesssim 500$ for advanced LIGO) across the whole
Universe.

Figure~\ref{fig:SNRLISA} shows similar information but for the planned
LISA mission. Since the horizon distance can well exceed the whole
Hubble horizon, the figure reports the averaged and maximum SNR
$\rho=\rho(a, M)$ for sources at $d=6.4\,\mathrm{Gpc}\
(z=1)$. Many of the considerations made above hold also for the LISA
detector, and it is interesting to note that for sufficiently high and
aligned spins (\ie $a\gtrsim 0.8$), the SNR is $\gtrsim {\cal O}(10)$
already with binaries having masses $\gtrsim {\rm few}\times
10^3\,M_{\odot}$.

\begin{figure*}
 \begin{center}
   \scalebox{0.425}{\includegraphics{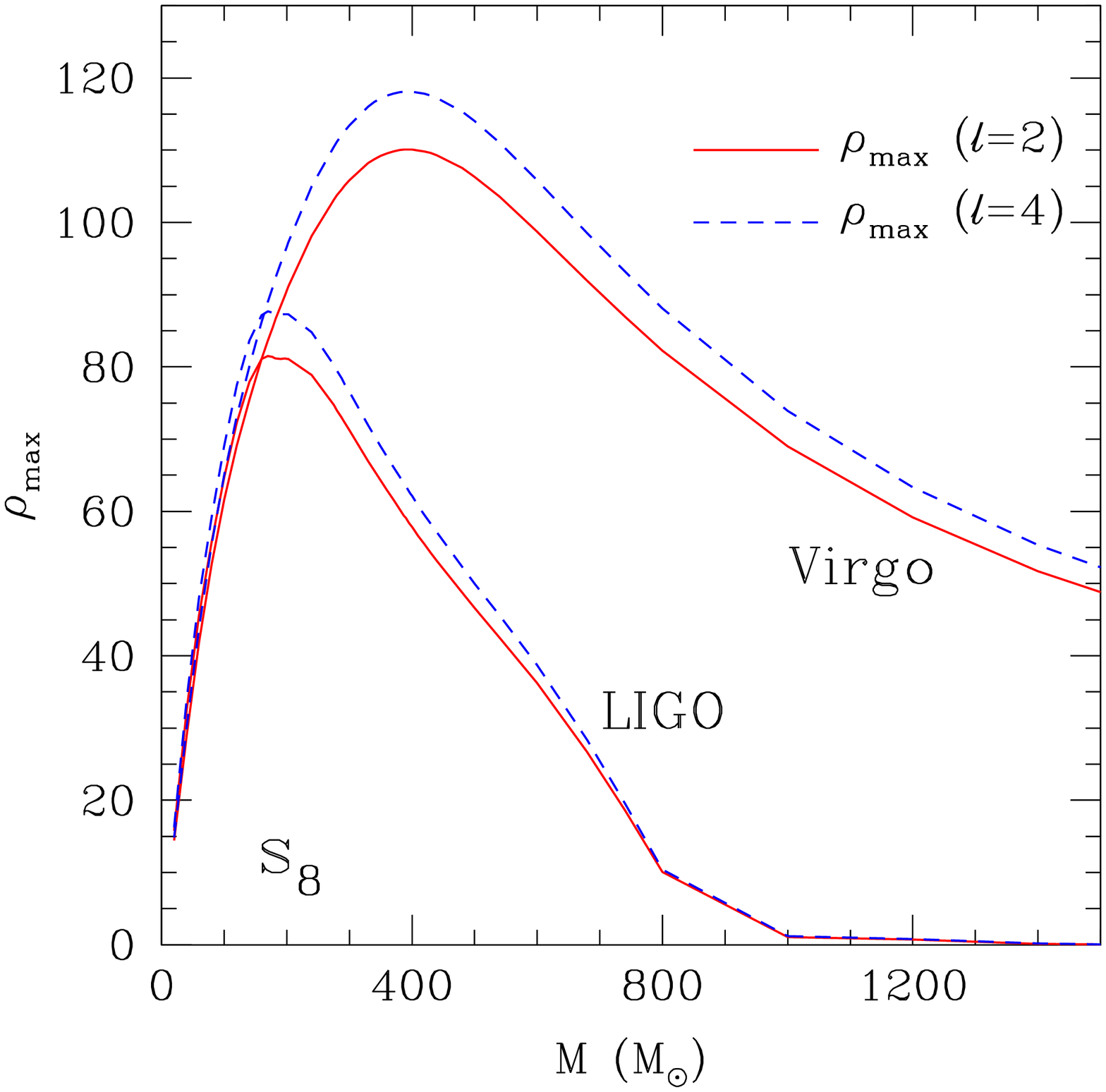}}
   \hskip 0.5cm
   \scalebox{0.425}{\includegraphics{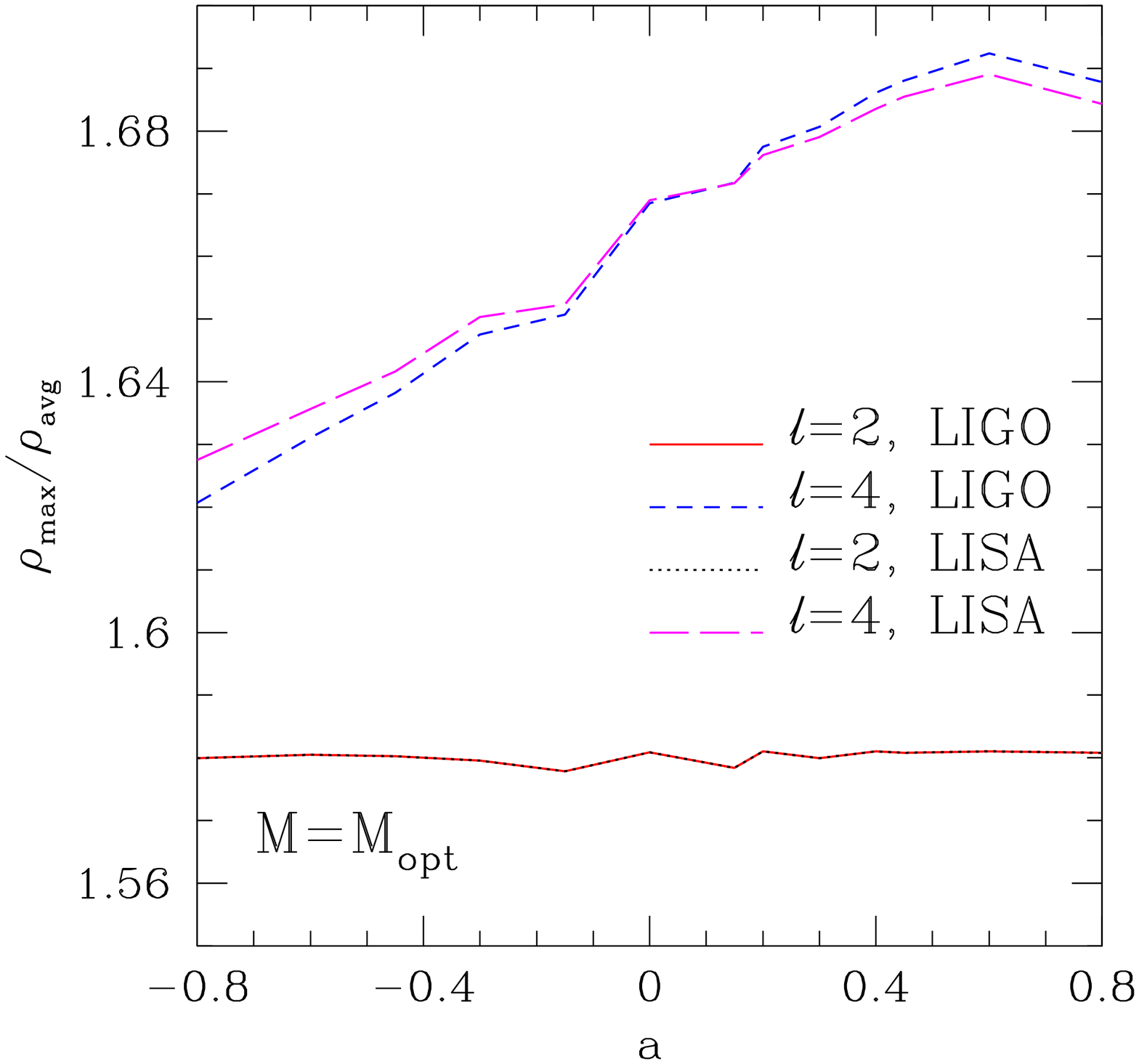}}
 \end{center}
 \caption{\textit{Left panel:} maximum SNR $\rho_{\rm max}$ as a
   function of the mass for the highly spinning model $s_8$ and for
   the present detectors LIGO and Virgo. Different lines refer to the
   SNRs computed using only the $\ell=2$ multipoles (continuous line),
   or up to the $\ell=4$ multipoles (dashed line). \textit{Right
     panel:} ratio between maximum and averaged SNR $\rho$ as a
   function of the spins $a_1=a_2$ for $M=200\,M_{\odot}$
   ($M=3.53\times 10^6\,M_{\odot}$) by including modes up to $\ell=2$
   and $\ell=4$ for LIGO (LISA). In contrast to the case $\ell=2$, the
   $\ell=4$-curve is not constant but depends on the initial spins
   $a_1, a_2$}
 \label{fig:max-avg}
\end{figure*}

Finally, the most salient information of Figs.~\ref{fig:SNRLIGO}
and~\ref{fig:SNRLISA} is collected in Table~\ref{tablefour} which
reports the properties of the ``optimal'' aligned binaries for the
different detectors. More specifically, the Table reports in its
different rows the optimal total aligned spin $a$, the optimal total
mass in solar masses, the optimal maximum $\rho$ and average
$\rho_{\rm avg}$ SNRs, the optimal horizon distance
$d_H$ (expressed in $\mpc$ and with $H^{-1}$ being the Hubble radius),
the optimal relative event rate $R$, and the glueing frequency $f_{\rm
  glue}$ for the optimal binary.  The masses have been sampled with an
accuracy of $2.5\,M_{\odot}$ for the ground-based detectors and of
$2.5\times 10^4\,M_{\odot}$ for LISA.

\begin{table}
\caption{\label{tablefour} Properties of the ``optimal'' aligned
  binaries for the different detectors. Shown in the different rows
  are the optimal total aligned spin $a$, the optimal total mass in
  solar masses, the optimal maximum $\rho_{\rm max}$ and average
  $\rho_{\rm avg}$ SNRs, the optimal horizon distance $d_H$ (expressed
  in $\mpc$ and where $cH^{-1}$ is the Hubble radius), the lower bound
  for the optimal relative event rate $R$, and the glueing frequency
  $f_{\rm glue}$ for the optimal binary. The masses have been sampled
  with an accuracy of $2.5\,M_{\odot}$ for the ground-based detectors
  and of $2.5\times 10^4\,M_{\odot}$ for LISA.}
\vspace{0.1cm}
\begin{ruledtabular}
\begin{tabular}{|l|r|r|r|r|r|r|}
\hline
                        & LIGO    & eLIGO    & AdLIGO  & Virgo  & AdVirgo  &  LISA \\
\hline
$a$                     & $0.8$   & $0.8$   & $0.8$   & $0.8$    & $0.8$   & $0.8$ \\
$M_{\rm opt}~(M_{\odot})$   & $197$   & $180$   & $290$   & $395$    & $390$   & $5.35\times 10^6$ \\
$\rho_{\rm max}$          & $87$    & $175$   & $1667$  & $118$    & $1591$  & $2.91\times 10^{6}$ \\
$\rho_{\rm avg}$          & $52$    & $104$   & $991$   & $70$     & $944$   & $1.77\times 10^{6}$ \\
$d_H~(\mpc)$            & $1091$  & $2190$  & $> cH^{-1}$ & $1476$ & $> cH^{-1}$ & $> cH^{-1}$ \\
$R$                     & $18$    & $17$    & $16$     & $16$     & $17$    & $26$ \\
$f_{\rm glue}~({\rm Hz})$ & $27.48$ & $30.51$ & $18.71$  & $13.74$ & $13.91$ & $1.0\times 10^{-3}$\\
\hline
\end{tabular} \\
\end{ruledtabular}
\end{table}

\subsection{Influence of higher $\ell$-modes}

As discussed in Sect.~\ref{sec:SNR}, it is interesting to consider the
impact that higher-order modes have on the SNR of equal-mass aligned
binaries and some representative examples of this impact is shown in
Fig.~\ref{fig:max-avg}. The left panel of this figure, in particular
shows the maximum SNR $\rho_{\rm max}$ as a function of the mass for
the highly spinning model $s_8$ and for the present detectors LIGO and
Virgo. Different lines refer to the SNRs computed using only the
$\ell=2$ multipoles (continuous line), or up to the $\ell=4$
multipoles (dashed line). Clearly, the contribution of the higher
modes is most important near the optimal mass (\ie $M\sim
200\,M_{\odot}$ for LIGO and $M\sim 400\,M_{\odot}$ for Virgo) but
this is also non-negligible for larger masses, where it can produce an
increase of $\sim 8\%$ in SNR in a detector such as Virgo.

The right panel of Fig.~\ref{fig:max-avg}, on the other hand, shows
the ratio between maximum and averaged SNR as a function of the total
projected spin $a$ for a binary of $M=200\,M_{\odot}$ ($5.35 \times
10^6\,M_{\odot}$) and the LIGO (LISA) detector.  As mentioned in
Sect.~\ref{sec:SNR}, this ratio is not expressed by a simple algebraic
expression [\cf equation~\eqref{eq:SNRR}], but needs to be determined
numerically.  Interestingly, this ratio is not constant but increases
by $\sim 10\%$ for larger total projected spins, underlining the
importance of higher-order contributions as the initial spins
increase. Overall, therefore, Fig~\ref{fig:max-avg} provides the
answer to question \textit{(iii)} in the Introduction.

\subsection{Match between different models}
\label{sec:match}

A quantity providing a wealth of information is the match between the
amplitudes of the waveforms from two different binaries, so as to
quantify the differences in the gravitational-wave signal relative to
some reference models. The match between two waveforms $h_1(t)$ and
$h_2(t)$ (or a template and a waveform) can be calculated via the
weighted scalar product in frequency space between two given waveforms
\begin{equation}
\langle h_1 | h_2 \rangle = 4 \Re \int_0^\infty df \frac{\tilde{h}_1(f) \tilde{h}_2^*(f)}{S_h(f)}\,,
\end{equation}
where $\tilde{h}_1(f)$ is the power spectral density of $h_1(t)$, the
asterisk indicates a complex conjugate, and $S_h(f)$ is the noise power
spectral density of a given detector.  The overlap is then simply
given by the normalized scalar product
\begin{equation}
\mathcal{O}[h_1, h_2] = \frac{\langle h_1 | h_2 \rangle}{\sqrt{\langle h_1 | h_1 \rangle \langle h_2 | h_2 \rangle}}.
\end{equation}

Two parameters need to be taken into account when computing the
overlap. The first one is the ``time of arrival'' $t_{\rm A}$
corresponding to an offset in the Fourier-transform of the signal
$\exp{\left[{\rm i}\omega(t-t_{\rm A})\right]}$. The second one is the
``initial phase'' $\Phi$ of the orbital motion when it enters the
detector band.

For both of these parameters the overlap should be maximized. We
have considered two possible ways of doing this. The first approach
involves the \textit{best} match, which gives an upper bound by
maximizing over both of the phases of each waveform
\begin{equation}
\label{eq:best}
{\cal M}_{\rm best}\equiv \max_{t_{\rm A}}\max_{\Phi_1}\max_{\Phi_2}
\{\mathcal{O}[h_1, h_2]\}\,.
\end{equation}
The second way, instead, involves the \textit{minimax} match, and is
obtained by maximizing over the phase of one waveform but minimizing
over the phase of the other
\begin{equation}
\label{eq:minmax}
{\cal M}_{\rm minimax}\equiv \max_{t_{\rm A}}\min_{\Phi_2}\max_{\Phi_1}
\{\mathcal{O}[h_1, h_2]\}\,,
\end{equation}
and thus represents a ``worst-case'' scenario since it gives lower
matches although one is maximizing over the template phase.  More
details on the maximization procedure can be found in
\cite{Damour_T:98, Vaishnav:2007nm}.  Note that all the matches
computed hereafter refer to the numerical-relativity part of the
waveform only.

A sensible way, if not the most sensible way, of evaluating
expressions~\eqref{eq:best} and ~\eqref{eq:minmax} is to use the
binary $s_0$, the nonspinning binary, as a reference and to compute
the overlap with the binaries at representative locations in the spin
diagram, \eg at the corners for $s_0-s_8$, $s_0-u_8$, $s_0-s_{-8}$, or
along the main diagonal, \eg $s_{-8}-s_8$. In this way we can assess
whether the waveform produced by a nonspinning binary can be used to
detect also spinning binaries and how much the overlap is decreased in
this case.

\begin{figure}
 \begin{center}
   \scalebox{0.425}{\includegraphics{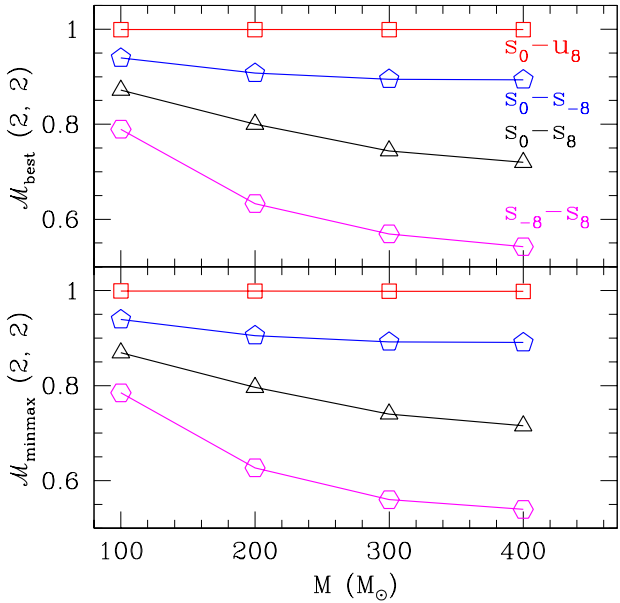}}
 \end{center}
 \caption{Best and minmax match as a function of mass for a
   waveform containing only the $\ell=2,m=2$ contribution and
   referring to the LIGO detector. Very similar behaviors can be
   shown also for the other detectors.}
\label{fig:match_1}
\end{figure}

This is shown in Fig.~\ref{fig:match_1}, which reports the best and
minmax matches as a function of mass for a waveform containing only
the $\ell=2,m=2$ contribution and refers to the LIGO detector.
Different lines show the match computed between $s_0$ and other
representative binaries, and show the remarkable similarity between
the waveforms of binaries having a zero total spin. This is shown by
the $s_0-u_8$ match, which is essentially very close to $1$ for all
the masses considered (\cf also Table~\ref{tablethree}). This result
extends to all the other measured quantities, such as the radiated
energy or angular momentum, and is not particularly
surprising. Indeed, it was already discussed
by~\cite{Vaishnav:2007nm}, although the investigation in that case was
restricted to what is here the $u$-sequence. In addition, the
equivalence between nonspinning binaries and binaries with equal and
opposite spins has been exploited in the derivation of expressions for
the final spin presented in a series of
works~\cite{Rezzolla-etal-2007, Rezzolla-etal-2007b,
  Rezzolla-etal-2007c,Barausse:2009uz}. The results of
Fig.~\ref{fig:match_1} and Table~\ref{tablethree} are therefore a
simple example, although probably not the only possible one, of a well
defined region of the space of initial configurations (\ie those of
binaries with equal masses and opposite spins) which can be mapped to
an almost degenerate region (\ie essentially to a single point) in the
space of templates. This is the answer to question \textit{(iv)} in
the Introduction and clearly represents a serious obstacle towards a
proper estimate of physical parameters of the binaries that may be
removed, at least in part, only if the waveform is measured with a
sufficiently high SNR. A proper discussion of this problem, as well as
the determination of other degenerate patches in the space of
templates, will be the subject of future work.

An equally remarkable result, presented in Fig.~\ref{fig:match_1}, is
that the overlap is also very high between the nonspinning binary and
the binary with equal and antialigned spins, $s_0-s_{-8}$; also in
this case, in fact, the best match is ${\cal M}_{\rm best} \gtrsim
0.9$ for the range of masses that is relevant here. Slightly smaller
and decreasing with increasing masses are the best matches computed
when comparing the nonspinning binary with the binary of parallel
and aligned spins, so that ${\cal M}_{\rm best} \sim 0.8$, but only
for very large masses. The waveforms appear clearly different (\ie
with ${\cal M}_{\rm best} \lesssim 0.6$) only when comparing the
binaries along the main diagonal of the spin diagram, for
$s_8-s_{-8}$, although even in this rather extreme case the
differences tend to become smaller for smaller masses. Overall, this
result underlines that even simple waveforms, such as those relative
to nonspinning binaries, will be effective enough to provide a
detection for most configurations of equal-mass and
aligned/antialigned binaries.

\begin{figure}
 \begin{center}
   \scalebox{0.425}{\includegraphics{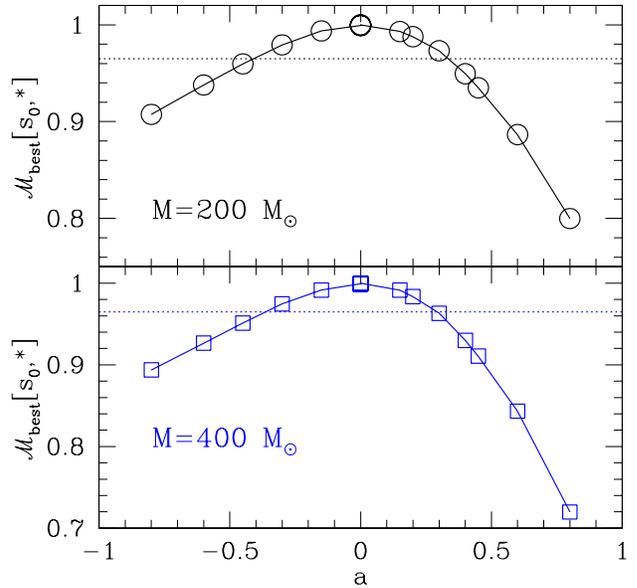}}
 \end{center}
 \caption{Best match as a function of the total projected spin $a$ for
   a waveform containing only the $\ell=2,m=2$ contribution. The
   top/lower panels refers to a binary with a total mass
   ($200/400\,M_{\odot}$) which are close to the optimal ones for the
   LIGO/Virgo or advanced detectors, respectively. In both panels the
   dotted line shows the minimum best match ($0.965$) needed for a
   detection. While the data have been computed for the LIGO detector,
   very similar behaviors can be shown also for the other detectors.}
\label{fig:match_2}
\end{figure}

A different way to assess ``how different'' the waveforms are across
all of the equal-mass aligned/antialigned spins configurations
considered here is nicely summarized in Fig.~\ref{fig:match_2}, which
shows the best match as a function of the total projected spin $a$ for
waveforms containing only the $\ell=2,m=2$ contribution and referring
to the LIGO detector. The top panel, in particular, refers to a binary
with a total mass of $200\,M_{\odot}$ that is close to the optimal one
for the LIGO/Virgo detectors, while the bottom panel refers to a
binary with mass $400\,M_{\odot}$ and close to the optimal one for the
advanced LIGO/Virgo detectors (\cf Table~\ref{tablethree}). Besides
the remarkably smooth behaviour of ${\cal M}_{\rm best}$ across all
the values of $a$ considered, it is clear that the waveform from a
nonspinning binary can be extremely useful across the \textit{whole}
spin diagram and yield very large overlaps even for binaries with very
high spins. In both panels, in fact, the dotted line shows the minimum
best match (${\cal M}_{\rm best}=0.965$) needed for a
detection~\cite{Abbott:2005qm}.  This result is reassuring in light of
the fact that most of the searches in the detector data are made using
phenomenological waveforms based on nonspinning binaries.

For completeness, the results presented in Fig.~\ref{fig:match_1} (as
well as those in Fig.~\ref{fig:match_3}) are also reported in
Table~\ref{tablethree}, where the different columns show ${\cal
  M}_{\rm best}$ and ${\cal M}_{\rm minmax}$ and for waveforms
computed either using only the $\ell=2, m=2$ contribution (third and
fourth columns), only the $\ell=3, m=2$ contribution (fifth and sixth
columns), or all contributions up to $\ell=4$ (last two
columns). Interestingly, the matches among the high-order modes, \eg
$(s_0)_{\ell=3, m=2}-(u_8)_{\ell=3, m=2}$, is systematically higher
than those of the lower ones and remains true even for higher modes
beyond $\ell=3, m=2$ which, however, we do not report here. This
indicates that in order to do high-precision parameter estimation by
including higher modes it is also important that these modes are
accurately resolved, so that they can be clearly distinguished from
one another.

We generally expect the match to degrade when the waveforms are
computed by including higher-order modes (\eg up to $\ell=4$) and that
this degradation will become larger with increasing inclination
$\theta$. The most notable example is for the degeneracy along the
diagonal $a_1=-a_2$, which should be broken by the inclusion of
higher-order modes (We recall that these configurations lead to
different recoil velocities~\cite{Rezzolla-etal-2007} which can only
be produced by gravitational-wave contributions other than the leading
order $\ell=m=2$ mode). For this reason we have computed the
sky-averaged match of waveforms including modes up to $\ell=4$ (\ie
the ``complete'' waveforms) and the corresponding matches are reported
in the last two columns of Table~\ref{tablethree}. Similarly to what
was found in~\cite{Vaishnav:2007nm}, we measure a marked decreased in the
minmax match, but a much smaller decrease in the best match (the
latter was not considered in~\cite{Vaishnav:2007nm}). Although our
resolution should be marginally enough for us to detect such a
difference in the best match, we also believe that a much higher
accuracy is required to determine this with certainty. Note also that
the matches with complete waveforms along other directions,~\eg
$s_0-s_8$ or $s_0-s_{-8}$ do not decrease and this is simply due to
the very large mismatch we already have with the $\ell=2=m$-waveforms
(in these cases, in fact, the final black holes are considerably
different and hence the associated ringdowns are expected to be
different).

Finally, we note that although Figs~\ref{fig:match_1}
and~\ref{fig:match_2} show data computed for the LIGO detector, very
similar behaviors can be shown also for the other detectors.

\begin{table*}
\caption{\label{tablethree} Best and minmax matches as computed for
  the LIGO detector for binaries with different spins in the spin
  diagram. Different columns show ${\cal M}_{\rm best}$ and ${\cal
    M}_{\rm minmax}$ for waveforms computed either using only the
  $\ell=2, m=2$ contribution (third and fourth columns), only the
  $\ell=3, m=2$ contribution (fifth and sixth columns), or the
  sky-averaged contributions of all modes up to $\ell=4$ (last two
  columns). Finally the last eight rows show the matches at different
  resolutions (\ie $\Delta x/M=0.024, 0.020, 0.018$ or low, medium and
  high, respectively) for the binary $r_0$.}
\vspace{0.1cm}
\begin{ruledtabular}
\begin{tabular}{|l|c|c|c|c|c|c|c|}
\hline
~				&
$M/M_{\odot}$	   	        &
${\cal M}_{\rm best}$ 		&
${\cal M}_{\rm minmax}$ 		&
${\cal M}_{\rm best}$ 		&
${\cal M}_{\rm minmax}$ 		&
${\cal M}_{\rm best}$ 		&
${\cal M}_{\rm minmax}$ 		\\
~				&
~	   	                &
only $\ell = 2, m=2$ 		&
only $\ell = 2, m=2$ 		&
only $\ell = 3, m=2$ 		&
only $\ell = 3, m=2$            &
avg. up to $\ell = 4$ 		&
avg. up to $\ell = 4$            \\
\hline
$s_0-s_8$   & $100$ & $0.87182$ & $0.86914$ & $0.87802$ & $0.85061$ & $0.86337$ & $0.83272$ \\
            & $200$ & $0.79987$ & $0.79642$ & $0.82533$ & $0.80236$ & $0.80070$ & $0.75679$ \\
            & $300$ & $0.74394$ & $0.74026$ & $0.82570$ & $0.78819$ & $0.74785$ & $0.71139$ \\
            & $400$ & $0.71981$ & $0.71568$ & $0.84074$ & $0.81285$ & $0.72345$ & $0.69019$ \\
\hline                                                                                      
$s_0-u_8$   & $100$ & $0.99926$ & $0.99914$ & $0.99497$ & $0.97411$ & $0.99673$ & $0.95443$ \\
            & $200$ & $0.99928$ & $0.99906$ & $0.99372$ & $0.95193$ & $0.99483$ & $0.95919$ \\
            & $300$ & $0.99923$ & $0.99870$ & $0.99189$ & $0.93888$ & $0.99251$ & $0.96105$ \\
            & $400$ & $0.99919$ & $0.99822$ & $0.99147$ & $0.93493$ & $0.99110$ & $0.96054$ \\
\hline                                                                                      
$s_0-s_{-8}$ & $100$ & $0.93942$ & $0.93907$ & $0.95717$ & $0.94843$ & $0.93695$ & $0.92143$ \\
            & $200$ & $0.90746$ & $0.90536$ & $0.95647$ & $0.94521$  & $0.89646$ & $0.88041$ \\
            & $300$ & $0.89491$ & $0.89197$ & $0.95015$ & $0.93814$  & $0.87303$ & $0.84960$ \\
            & $400$ & $0.89369$ & $0.89065$ & $0.94806$ & $0.93550$  & $0.85492$ & $0.82103$ \\
\hline                                                                                      
$s_{-8}-s_8$ & $100$ & $0.78948$ & $0.78493$ & $0.87041$ & $0.85222$ & $0.78310$ & $0.74895$ \\
            & $200$ & $0.63309$ & $0.62703$ & $0.90722$ & $0.88543$ &  $0.63456$ & $0.59426$ \\
            & $300$ & $0.56934$ & $0.56008$ & $0.90322$ & $0.88869$ &  $0.56941$ & $0.52170$ \\
            & $400$ & $0.54235$ & $0.53960$ & $0.91199$ & $0.89848$ &  $0.55470$ & $0.49338$ \\
\hline                                                                                      
$s_{-8}-u_8$ & $100$ & $0.94250$ & $0.94187$ & $0.96299$ & $0.94669$ & $0.93897$ & $0.89017$ \\
            & $200$ & $0.91444$ & $0.91229$ & $0.96316$ & $0.93068$  & $0.90315$ & $0.85958$ \\
            & $300$ & $0.90188$ & $0.89885$ & $0.95486$ & $0.91256$  & $0.87846$ & $0.83428$ \\
            & $400$ & $0.89772$ & $0.89492$ & $0.95132$ & $0.90583$  & $0.85870$ & $0.80907$ \\
\hline                                                                                      
$s_8-u_8$   & $100$ & $0.87127$ & $0.86817$ & $0.87656$ & $0.84229$ & $0.85866$ & $0.80969$ \\
            & $200$ & $0.79750$ & $0.79477$ & $0.83582$ & $0.81476$ & $0.79074$ & $0.73526$ \\
            & $300$ & $0.74063$ & $0.73884$ & $0.83897$ & $0.80378$ & $0.73616$ & $0.68774$ \\
            & $400$ & $0.71798$ & $0.71343$ & $0.84955$ & $0.81925$ & $0.71203$ & $0.66611$ \\
\hline                                                                                      
\hline                                                                                      
$r_0$
	    & $100$ & $0.99979$ & $0.99970$ & $0.99495$ & $0.98812$ & $0.99855$ & $0.99463$ \\
$(0.024, 0.020)$                                                                  
            & $200$ & $0.99963$ & $0.99929$ & $0.99133$ & $0.97100$ & $0.99633$ & $0.98800$ \\
            & $300$ & $0.99943$ & $0.99894$ & $0.98752$ & $0.95775$ & $0.99379$ & $0.98152$ \\
            & $400$ & $0.99924$ & $0.99868$ & $0.98630$ & $0.95317$ & $0.99209$ & $0.97683$ \\
\hline                                                                                      
$r_0$
	    & $100$ & $0.99990$ & $0.99989$ & $0.99873$ & $0.99299$ & $0.99881$ & $0.99639$ \\
$(0.020, 0.018)$                                                                  
            & $200$ & $0.99980$ & $0.99970$ & $0.99806$ & $0.98074$ & $0.99705$ & $0.98952$ \\
            & $300$ & $0.99956$ & $0.99924$ & $0.99707$ & $0.97238$ & $0.99497$ & $0.98070$ \\
            & $400$ & $0.99935$ & $0.99866$ & $0.99666$ & $0.97017$ & $0.99320$ & $0.97429$ \\
\hline
\end{tabular} 
\end{ruledtabular}
\end{table*}

\subsection{Accuracy of NR waveform amplitudes}
\label{sec:amplitude_accuracy}

\begin{figure}
 \begin{center}
   \scalebox{0.425}{\includegraphics{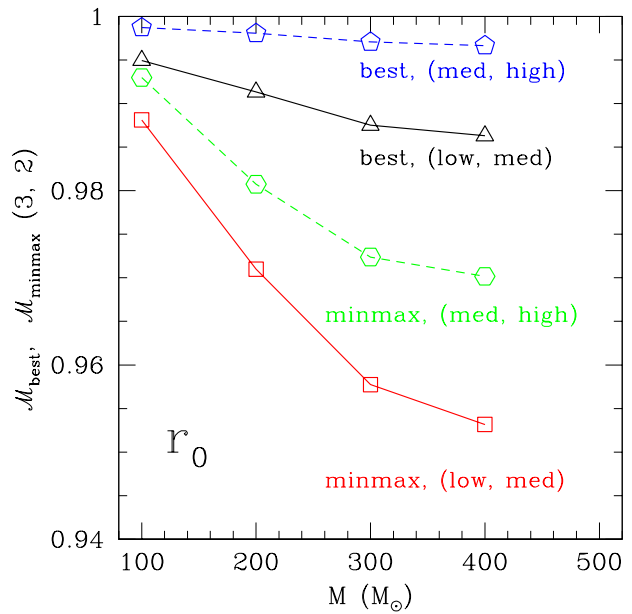}}
 \end{center}
 \caption{As in Fig.~\ref{fig:match_2} but now different lines
   represent the matches obtained when comparing the numerical
   waveforms of the binary $r_0$ computed at different
   resolutions. The matches are computed for the LIGO detector, but
   very similar behaviors can be shown also for the other detectors.}
\label{fig:match_3}
\end{figure}

A reasonable concern that can be raised when looking the very high
matches between the waveforms in the $u$-sequence is that these are
simply the result of insufficient resolution. In other words, the
waveforms may appear similar simply because our resolution is not
sufficient to pick-up the differences. To address this concern we have
computed the overlap among the waveforms obtained at three different
resolutions and for a representative binary with nonzero spins, \ie
$r_0$. Clearly, a low match in this case would be an indication that
our results are very sensitive to the numerical resolution and hence
the conclusions drawn on the degeneracy of the space of templates
would be incorrect.

The results of this validation are presented in Fig.~\ref{fig:match_3}
and are reported in the last eight rows of
Table~\ref{tablethree}. More specifically, shown with different lines
in Fig.~\ref{fig:match_3} are the matches obtained when comparing the
numerical waveforms of the binary $r_0$ computed at low resolution
($\Delta x/M=0.024$) and medium resolution ($\Delta x/M=0.020$, which
is also the standard one), as well as at a medium and high resolution
($\Delta x/M=0.018$). The matches are computed considering only the
$\ell=2,m=2$ mode and for the LIGO detector, but very similar
behaviors can be shown also for higher modes or for the other
detectors.

Overall, the results reported in Fig.~\ref{fig:match_3} and in
Table~\ref{tablethree} show that ${\cal M}_{\rm best, minmax}[\Delta
  x_1, \Delta x_2] > {\cal M}_{\rm best, minmax}[h_1, h_2]$, \ie that
the differences we measure in the overlaps among two different
waveforms $h_1$ and $h_2$ are always larger than the differences we
are able to measure at two different resolutions $\Delta x_1$ and
$\Delta x_2$. In other words, the differences in the waveforms across
the spin diagram are always larger than our numerical errors, even
along the degenerate $u$-sequence (of course, as we have a convergent
numerical code, the match between medium and low resolution is worse
than the match between medium and high resolution). It is also worth
mentioning that as long as the dominant $\ell=2, m=2$ mode is
considered, the differences in the matches are well within the margin
of error for numerical-relativity simulations of black hole
binaries. A recent work has in fact estimated that the differences in
the waveforms produced by distinct codes is ${\cal M}_{\rm
  mismatch}=1-{\cal M}\approx10^{-4}$ for the last $\approx 1000 M$ of
the dominant mode of non-spinning equal mass
coalescence~\cite{Hannam:2009hh}. Since the next higher mode $\ell=3,
m=2$ starts to suffer from numerical noise, it does not yield the same
high agreement, and the differences between best and minimax match
show a larger deviation.

As a final comment on the accuracy of our waveforms, we note that the
error made by using waveforms extracted at a finite radius, and not
extrapolated at spatial infinity is well within the error budget of
our estimates. We have validated this by comparing the waveforms
extracted at a finite radius against the waveforms computed at future
null infinity, via a newly developed Cauchy-characteristic
code~\cite{Reisswig:2009us}. In the case of the
nonspinning configuration $s_0$ we have found an error in the
calculated SNR of less than $1.0\%$ (details on this comparison can be
found in Appendix~\ref{sec:appendix_b}).

\section{Fitting formulas}
\label{sec:FF}

In what follows we provide some simple analytic representation of most
of the results presented in the previous Sections and, in particular,
we give a brief discussion of fitting expressions that can be derived
to express the SNR for an optimal mass and the energy radiated in
gravitational waves.

\subsection{SNR}
\label{sec:fitSNR}

As discussed in Sect.~\ref{sec:SNR}, the maximum SNR depends on
several factors, most notably on the two initial spins, the total mass
of the system and, although more weakly, on the number of multipoles
included in the waveforms. The resulting functional dependencies when
one degree of freedom is suppressed and the SNRs are presented in
terms of the total projected spin are shown in
Figs.~\ref{fig:SNRLIGO},~\ref{fig:SNRLISA} and are clearly too
cumbersome to be described analytically (although still possible).

\begin{figure}
 \begin{center}
   \scalebox{0.425}{\includegraphics[angle=-0]{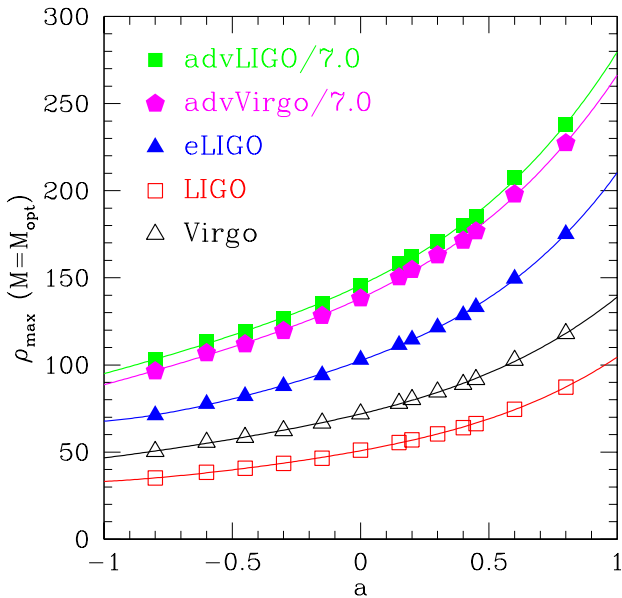}}
 \end{center}
 \caption{Different symbols show the numerically computed values of
   $\rho_{\rm max}(a, M_{\rm opt})$ for the different detectors and
   represent therefore the cross section along the optimal mass of
   Figs.~\ref{fig:SNRLIGO} and~\ref{fig:SNRLISA}. Note that the SNR
   for the advanced detectors have been divided by $7$ to make them
   fit onto the same scale.}
\label{fig:rhomax_vs_a}
\end{figure}

However, most of the complex functional dependence can still be
captured when concentrating on the best case scenario, and hence on the
SNRs relative to the optimal mass $M_{\rm opt}$. The behaviour of the
SNR in this case is shown in Fig.~\ref{fig:rhomax_vs_a}, where the
different symbols show the numerically computed values of $\rho_{\rm
  max}(a, M_{\rm opt})$ for the different detectors. Stated
differently, Fig.~\ref{fig:rhomax_vs_a} represents the cross section
along the optimal mass of Figs.~\ref{fig:SNRLIGO}
and~\ref{fig:SNRLISA} (note that the SNR for the advanced detectors
have been divided by $7$ to make them fit onto the same scale).

Clearly, the behaviour of the SNR in this case is sufficiently simple
that it can be represented with a simple quartic polynomial of the
type
\begin{equation}
\rho_{\rm max}(a;\, \ell \leq 4, M=M_{\rm opt})  =
	\sum^4_{n=0} k_n a^n   \,,
\label{af_1}
\end{equation}
whose coefficients $k_n$ are reported in Table~\ref{tablefive} for
the five detectors considered.

These results address therefore question \textit{(ii)} formulated in
the Introduction. More specifically, when considering the optimal
mass, the ratio of the SNRs for maximally antialigned spinning
binaries to maximally and aligned spinning binaries, \ie $\rho_{\rm
  max}(a=1)/\rho_{\rm max}(a=-1)$ is $\sim 3$ for both the LIGO and
Virgo detectors.  This ratio is also preserved when considering the
advanced LIGO and Virgo detectors. Because the event rate scales like
the cube of the SNR [\cf expressions~\eqref{eq:SNRR}-\eqref{eq:R}], an
increase of a factor $\sim 3$ in the SNR of binaries with $a=-1$ and
$a=1$ will translate into an increase of a factor $\sim 27$ in the
event rate. It is therefore likely that many of the binaries observed
will have high spins and aligned with the orbital angular
momentum. This will be particularly true in the case of LISA if the
prediction that the spins of supermassive black holes are aligned
with the orbital angular momentum will hold~\cite{Bogdanovic:2007hp}.

\subsection{Radiated Energy}
\label{sec:fitErad}

While the SNR is effectively a measure of the amount of energy
released during the inspiral, it also incorporates information on the
properties of the detectors and is not therefore an absolute measure
of the efficiency of the gravitational-wave emission process. This
information can have a number of important astrophysical applications,
and in particular it can be used to study the effect the merger has on
the dynamics of the circumbinary disk accreting onto the binary when
this is massive (see~\cite{Bode_Phinney_2007} for the first suggestion
and~\cite{Megevand2009} for a recent nonlinear study).

\begin{table}
\caption{\label{tablefive} Fitting coefficients for the maximum SNR
  computed for the optimal mass [\cf eq.~\eqref{af_1}]. The different
  rows refer to the various detectors and have been computed including
  all modes up to $\ell=4$.}
\vspace{0.1cm}
\begin{ruledtabular}
\begin{tabular}{|l|r|r|r|r|r|}
\hline
detector & $k_0$     & $k_1$    & $k_2$   & $k_3$   & $k_4$ \\
\hline
LIGO     & $50.76$   & $27.11$  & $13.43$  & $8.58$   & $4.63$  \\
eLIGO    & $102.45$  & $53.63$  & $25.33$  & $17.67$  & $11.26$ \\
AdLIGO   & $1020.42$ & $492.25$ & $243.60$ & $153.84$ & $46.99$ \\
Virgo    & $71.86$   & $35.23$  & $17.140$ & $10.92$  & $3.789$ \\
AdVirgo  & $968.08$  & $481.52$ & $236.45$ & $140.69$ & $37.91$ \\
\end{tabular} \\
\end{ruledtabular}
\end{table}

In this Section we present a simple formula to compute the amount of
energy released and express it only in terms of the initial spins. Our
formula is restricted to aligned binaries and is therefore not as
generic as the one recently presented in~\cite{Lousto:2009mf}, which
however also requires the determination of a larger set of
coefficients, some of which have uncertainties of $\sim 100\%$. As we
will show below, the two expressions yield results in reasonably good
agreement, at least in the part of the parameter space we investigate.

In practice, the expression for the radiated energy $E_{\rm rad}$ is
derived by combining a fit to the numerical data for the binaries at
an initial and finite separation $D=8\,M$~\footnote{Note that for the
  binary $s_0$, we use an initial separation of $D=10M$. In order to
  obtain the radiated energy obtained during a simulation starting
  from an initial separation of $D=8M$, we only need to recalculate
  the initial ADM mass of the spacetime for this initial separation.
  The final mass of the remnant is in fact the same.} (we refer to
this energy as to $E^{\rm NR}_{\rm rad}$), with the estimate of the
energy released from the binary when it goes from an infinite
separation down to $D$ (we refer to this energy as $E^{\rm PN}_{\rm
  rad}$), \ie
\begin{equation}
E_{\rm rad}=E^{\rm NR}_{\rm rad} + E^{\rm PN}_{\rm rad} =
M_{\rm ADM}-M_{\rm fin} + E^{\rm PN}_{\rm rad} \,,
\end{equation}
where $M_{\rm ADM}$ is the initial ADM mass as measured at spatial
infinity of the binary with separation $D$, and $M_{\rm fin}$ the
Christodoulou mass of the final black hole~\footnote{Note that $M_{\rm
    ADM}+ E^{\rm PN}_{\rm rad}$ is effectively the mass of the system
  when it has an infinite separation. This is approximately set to 1
  in most simulations but with a precision which is smaller than the
  one needed here.}. For the fit of the radiated energy during the
numerical evolution, $E^{\rm NR}_{\rm rad}$, we use the same symmetry
arguments first made in~\cite{Rezzolla-etal-2007} and then
successfully used
in~\cite{Rezzolla-etal-2007b,Rezzolla-etal-2007c,Barausse:2009uz} to
write a simple expression which is a Taylor expansion in terms of the
initial spins
\begin{eqnarray}
\frac{E^{\rm NR}_{\rm rad}(q=1,a_1,a_2)}{M} = p_0 + p_1 (a_1 +
  a_2) + p_2 (a_1 + a_2)^2 \,.\nonumber\\
\label{af_erad_1.5}
\end{eqnarray}
Fitting then the numerical data we
obtain the following values for the coefficients 
\begin{eqnarray}
&&p_0 = \frac{3.606 \pm 0.0271}{100}\,, \qquad 
	p_1 = \frac{1.493 \pm 0.0260}{100}\,, \nonumber \\
&&p_2 = \frac{0.489 \pm 0.0254}{100}\,.
\label{p0-p2}
\end{eqnarray}
where the reduced chi-squared is $\chi^2_{\rm red}=0.008$, and where
the largest error is in the 2nd-order coefficient but this is only
$\sim 5\%$. The different
coefficients~\eqref{p0-p2} can then be interpreted as the nonspinning
orbital contribution to the energy loss ($p_0$, which is the largest
and of $\sim 3.6\%$), the spin-orbit contribution ($p_1$, which is
$\lesssim 3.0\%$), and the spin-spin contribution ($p_2$, which is
$\lesssim 2.0\%$).  The relative error between the numerically
computed value of $E^{\rm NR}_{\rm rad}$ and the fitted one is
reported in the last column of Table~\ref{tableone}.

The PN expression for the energy radiated by the binary when going
from an infinite separation down to a finite one $r=d$, depends on the
total mass of the binary, the mass ratio and the spin components, \ie
$E^{\rm PN}_{\rm rad}=E^{\rm PN}_{\rm rad}(r, M, \nu, a_1, a_2)$,
which is the generalization to unequal masses of the energy expression
used in the definition of the TaylorT1 approximant in
ref.~\cite{Hannam:2007wf}. However, exploiting the fact that for
equal-mass binaries the PN radiated energy $E^{\rm PN}_{\rm rad}$
follows the same series expansion used for $E^{\rm NR}_{\rm rad}$, we
obtain for $M=1=q$
\begin{eqnarray}
&&\frac{E^{\rm PN}_{\rm rad}(a_1, a_2)}{M} = E^{{\rm PN}}_{\rm rad,0}\nonumber \\
&&\hskip 2.cm 
+ E^{{\rm PN}}_{\rm rad, 1} (a_1 + a_2) 
+ E^{{\rm PN}}_{\rm rad, 2} (a_1 + a_2)^2\,, \nonumber \\
\label{af_erad_2}
\end{eqnarray}
where the coefficients for $D=8\,M$ are given by
\begin{eqnarray}
\label{p0-p2_PN}
&&E^{{\rm PN}}_{\rm rad, 0} = \frac{6401}{524288} \simeq \frac{1.220}{100} \,,\nonumber \\
&&E^{{\rm PN}}_{\rm rad, 1} = \frac{985}{1048576 \sqrt{2}} \simeq \frac{0.0664}{100}\,,\nonumber \\ 
&&E^{{\rm PN}}_{\rm rad, 2} = -\frac{1}{32768} \simeq -\frac{0.00305}{100}\,.
\end{eqnarray}
A rapid inspection of the coefficients~\eqref{p0-p2_PN} is sufficient
to appreciate that the PN orbital contribution is only $\sim 33\%,$ the
one of the strong-field regime, but also that the spin-related PN
contributions are mostly negligible, being at most of $\sim 4\%$ as
produced in the last orbits.

We can now combine expressions~\eqref{af_erad_1.5}-\eqref{p0-p2} with
expressions~\eqref{af_erad_2}-\eqref{p0-p2_PN} and estimate that for
equal-mass binaries with aligned spins the energy radiated via
gravitational waves from infinity is
\begin{equation}
\frac{E_{\rm rad}(a_1,a_2)}{M} = {\tilde p}_0 + {\tilde p}_1 (a_1 + a_2) + 
\tilde{p}_2 (a_1 + a_2)^2\,,
\label{erad}
\end{equation}
where
\begin{eqnarray}
{\tilde p}_0 = \frac{4.826}{100}\,, \qquad 
{\tilde p}_1 = \frac{1.559}{100}\,,\qquad 
{\tilde p}_2 = \frac{0.485}{100}\,.
\label{erad_coeffs}
\end{eqnarray}
Of course these numbers are specific to equal-mass binaries and refer
to a situation in which the match between the PN evolution and the one
in the strong-field regime is made at a specific separation of
$D=8\,M$. However, we expect the results to depend only weakly on this
matching separation (as long as it is within a PN regime) and hence
that expressions~\eqref{erad} and~\eqref{erad_coeffs} are generically
valid at the precision we are considering them here, namely $\sim
5\%$. 

Using expression~\eqref{erad} a number of quantitative considerations
are possible. Firstly, the largest energy is clearly emitted by
equal-mass, maximally spinning binaries with spins parallel and
aligned with the orbital angular momentum at is $E_{\rm
  rad}(a=1)/M=9.9\%$. Excluding the astrophysically unlikely head-on
collision of two black holes moving near the speed of light (in which
case $E_{\rm rad} < 14\pm 3\%$~\cite{Sperhake:2008ga}), these binaries
are among the most efficient sources of energy in the
Universe. Secondly, equal-mass nonspinning binaries lose a
considerable fraction of their mass via radiation, with $E_{\rm
  rad}(a=0)/M=4.8\%$, while maximally spinning binaries with spins
parallel and antialigned with the orbital angular momentum have
$E_{\rm rad}(a=-1)/M=3.7\%$.

\begin{figure}
 \begin{center}
   \scalebox{0.425}{\includegraphics[angle=-0]{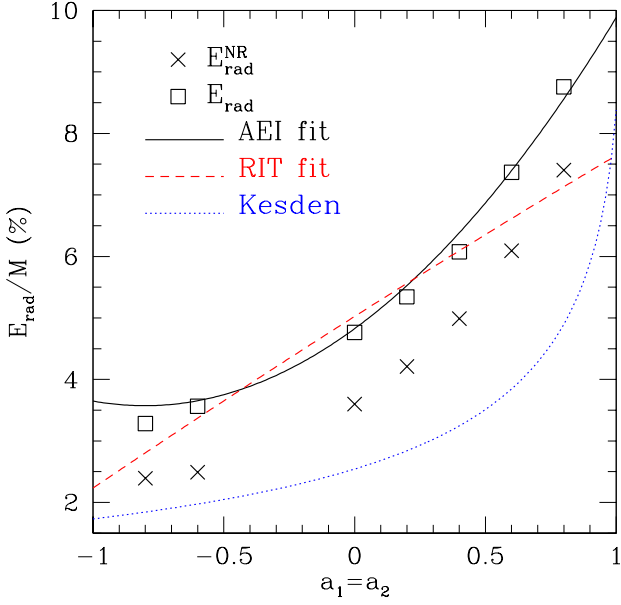}}
 \end{center}
 \caption{Energy radiated during the numerical calculation $E^{\rm
     NR}_{\rm rad}$ (crosses), the total radiated energy $E_{\rm
     rad}=E^{\rm NR}_{\rm rad}+E^{\rm PN}_{\rm rad}$ (squares) along
   the diagonal of the spin diagram, \ie for $a_1=a_2$. Shown as a
   continuous line is the analytic expressions given here (AEI fit),
   while the dashed line is the one suggested in
   ref.~\cite{Lousto:2009mf} (RIT fit). Note that the lines represent
   1-dimensional cuts of 2-dimensional surfaces and hence are not
   expected to fit well all points. Finally, indicated with a dotted
   line is the prediction for the radiated energy coming from the
   point-particle approach of~\cite{Buonanno:07b} and refined
   in~\cite{Kesden:2008}.}
\label{fig:Erad_vs_a}
\end{figure}

Note that expression~\eqref{erad} is not a strictly monotonic function
of the total spin and has a local minimum at $a_1=a_2 = -{\tilde
  p_1}/(4 {\tilde p_2}) \simeq -0.8$ rather than at $a_1=a_2=-1$, and
yields $E_{\rm rad}(a=-0.79)/M=3.6\%$ (\cf
Fig.~\ref{fig:Erad_vs_a}). Although rather shallow, we do not expect
such a local minimum. We therefore interpret it as an artifact of the
numerical error of our calculations (the difference between the energy
radiated at $a_1=a_2=-1$ and that at $a_1=a_2=-0.8$ is $\sim 2\%$ and
hence compatible with our overall error). Such a local minimum can be
removed by adding higher-order terms in expression~\eqref{af_erad_1.5}
(\eg up to 4th order in $a_1+a_2$) but these improvements are so small
that they do not justify the use of a more cumbersome expression. A
comparison between the numerical values and the fitting
expression~\ref{erad} is shown in Fig.~\ref{fig:Erad_vs_a}, where
crosses and squares represent the $E^{\rm NR}_{\rm rad}$ and $E_{\rm
  rad}$ respectively, along the diagonal of the spin-diagram (\ie for
$a_1=a_2$), while the continuous line refers to our fitting
expression. Note that such a line is a 1-dimensional cut of a
2-dimensional surface and hence it is not expected to exactly fit all
points.

As mentioned above, Lousto and collaborators~\cite{Lousto:2009mf} have
recently proposed a more general formula that should account for the
radiated energy in all of the relevant space of parameters, namely for
binaries with arbitrary mass ratio, spin orientation and
size. Restricting their expression to the specific subset of binaries
considered here corresponds to setting in their expression (2):
$E_B=E_E=0$, $\nu=1/4$ and $q=1$.  The resulting expression is then
\begin{eqnarray}
\frac{E^{\rm RIT}_{\rm rad}}{M} &=& \frac{1}{4}E_{\rm ISCO} + \frac{1}{16}E_2 + \frac{1}{64}E_3 \nonumber \\
  &+& \frac{1}{64} \left[ E_S (a_1+a_2) + E_A(a_1 + a_2)^2 \right. \nonumber \\
  &+& \left.E_D(a_1 - a_2)^2 \right]\,,
\label{eq:RIT-erad}
\end{eqnarray}
where the fitting coefficients have been determined to be $E_2=0.341
\pm 0.014$, $E_3=0.522 \pm 0.062$, $E_S=0.673 \pm 0.035$,
$E_A=-0.014\pm 0.021$, $E_D=-0.26 \pm 0.44$~\cite{Lousto:2009mf}, and
where
\begin{eqnarray}
E_{\rm ISCO} = \left(1-\frac{\sqrt{8}}{3}\right) + \frac{0.103803}{4} &&\nonumber \\
&&\hskip -4.0cm + \frac{1}{48\sqrt{3}}(a_1+a_2) + \frac{5}{648\sqrt{2}}(a_1 - a_2)^2\,.
\label{erad_rit}
\end{eqnarray}

After a bit of algebra we can rewrite~\eqref{erad_rit} as 
\begin{equation}
\frac{E^{\rm RIT}_{\rm rad}(a_1,a_2)}{M} = {\tilde q}_0 + 
{\tilde q}_1 (a_1 + a_2) + 
{\tilde q}_2 (a_1 + a_2)^2 +
{\tilde q}_3 (a_1 - a_2)^2\,,
\label{erad_RIT}
\end{equation}
where now
\begin{eqnarray}
&&{\tilde q}_0 = \frac{1}{4}\left(1 -\frac{\sqrt{8}}{3} + \frac{0.103803}{4}\right) + 
  \frac{E_2}{16} +  \frac{E_3}{64} \simeq \frac{5.025}{100} \,, \nonumber \\ 
&&{\tilde q}_1 = \frac{1}{192\sqrt{3}} + \frac{E_S}{64} \simeq \frac{1.352}{100}\,, \nonumber \\
&&{\tilde q}_2 = \frac{E_A}{64}\simeq -\frac{0.0219}{100}\,, \nonumber \\
&&{\tilde q}_3 = \frac{5}{2592\sqrt{2}} + \frac{E_D}{64} \simeq -\frac{0.270}{100}\,.
\label{erad_RIT_coeffs}
\end{eqnarray}

Comparing~\eqref{erad}-\eqref{erad_coeffs}
with~\eqref{erad_RIT}-\eqref{erad_RIT_coeffs} is now straightforward
and shows that: the reduced expression from~\cite{Lousto:2009mf}
has a second order contribution $\sim (a_1 - a_2)^2,$ which is absent
in our expression.  The remaining coefficients are rather
similar but not identical. This comparison is summarized in
Fig.~\ref{fig:Erad_vs_a}, where the dashed line corresponds to the
fitting proposed in ref.~\cite{Lousto:2009mf}. Note that the maximum
efficiency for maximally spinning black holes predicted by
expression~\eqref{erad_RIT} is $\sim 8\%$, but our estimate is
larger and $\sim 10\%$. Not reported in Fig.~\ref{fig:Erad_vs_a} is
the prediction made in ref.~\cite{Tichy:2008}, which is linear in the
total spin and very close to that coming from~\eqref{erad_RIT}.

While the two expressions provide very similar estimates for $-0.5
\lesssim a_1=a_2 \lesssim 0.4$, they also have predictions differing
by more than $\sim 20\%$ for highly spinning binaries. Because both
expressions come as a result of a number of simplifications and
assumptions, it is not easy to judge which one is the most accurate
one, if any. It is useful to bear in mind, however, that
expressions~\eqref{erad}-\eqref{erad_coeffs} have been obtained from a
``controlled'' set of simulations with small truncation errors and
therefore have coefficients with error-bars of the order of
$5\%$. Expressions~\eqref{erad_RIT}-\eqref{erad_RIT_coeffs}, on the
other hand, because coming from more extended formulas and thus
fitting a wider set of different simulations across many groups, have
error-bars that are intrinsically larger, as high as $100\%$. In view
of this, and of the fact that the coefficients are constant, the
simulations carried out here could be used for a new estimate of the
free coefficients $E_2, E_3, E_S,$ and $E_A$ in~\eqref{erad_RIT} (Note
that because in the expression for the radiated
energy~\eqref{af_erad_1.5} there is no need for a contribution
proportional to $(a_1 - a_2)^2$, it should be possible to set ${\tilde
  q}_3=0$ and obtain a numerical constraint for the presently
inaccurate coefficient $E_D$). Finally, indicated with a dotted line
in Fig.~\ref{fig:Erad_vs_a} is the prediction for the radiated energy
coming from the point-particle approach of~\cite{Buonanno:07b} and
refined in~\cite{Kesden:2008}.

Simulations involving aligned binaries with unequal masses will help
to settle this issue and provide an extension to our expression
~\eqref{erad}. This will be the subject of future work.

\section{Conclusions}
\label{sec:conclusions}

We have considered in detail the issue of the detectability of binary
system of black holes having equal masses and spins that are aligned
with the orbital angular momentum. Because these configurations do not
exhibit precession effects, they represent a natural ground to start
detailed studies of the influence of strong-field spin effects on
gravitational wave observations of coalescing binaries. Furthermore,
such systems are far from being unrealistic and may be the preferred
end-state of the inspiral of generic supermassive binary black-hole
systems. In view of this, we have computed the inspiral and merger of
a large set of binary systems of equal-mass black holes with spins
parallel to the orbital angular momentum but otherwise arbitrary. Our
attention is particularly focused on the gravitational-wave emission
so as to provide simple answers to basic questions such as what are
the  ``loudest'' and ``quietest'' configurations and what is the
difference in SNR between the two.

Overall we find that the SNR ratio increases with the projection of
the total black hole spin in the direction of the orbital momentum. In
addition, equal-spin binaries with maximum spin aligned with the
orbital angular momentum are more than ``three times as loud'' as the
corresponding binaries with anti-aligned spins, thus corresponding to
event rates up to $30$ times larger. On average these considerations
are only weakly dependent on the detectors, or on the number of
harmonics considered in constructing the signal.

We have also investigated whether these binaries can lead to a
degenerate patch in the space of templates.  We do this by computing
the mismatch between the different spinning configurations. Within our
numerical accuracy we have found that binaries with opposite spins
\mbox{$\boldsymbol{S}_1=-\boldsymbol{S}_2$} cannot be distinguished,
whereas binaries with spin $\boldsymbol{S}_1=\boldsymbol{S}_2$ have
clearly distinct gravitational-wave emissions. This result, which was
already discussed in the past~\cite{Vaishnav:2007nm}, may represent a
serious obstacle towards a proper estimate of the physical parameters
of binaries and will probably be removed only if the SNR is
sufficiently high.

Finally, we have derived a simple expression for the energy radiated
in gravitational waves, and find that the binaries always have
efficiencies $E_{\rm rad}/M \gtrsim 3.6\%$.  This can become as large
as $E_{\rm rad}/M \simeq 10\%$ for maximally spinning binaries with
spins aligned to the orbital angular momentum. These binaries are,
therefore, among the most efficient sources of energy in the Universe.

\acknowledgments

\noindent It is a pleasure to thank S. Babak, E. Barausse, M. Hannam, I. Hinder,
S. Hughes, B. Krishnan, L. Santamaria, B. Sathyaprakash and B. Schutz
for useful discussions and comments. Mathematica codes for 
post-Newtonian waveforms and waveform analysis have been developed
together with Mark Hannam. We thank E. Cuoco, S. Hild and
M. Punturo for providing the sensitivity curve of the advanced Virgo
detector. SH and DP have been supported as VESF fellows of the
European Gravitational Observatory (EGO). Additional support comes
from the DAAD grant D/07/13385, grant FPA-2007-60220 from the
Ministerio de Educaci\'on y Ciencia (Spain), and by DFG grant
SFB/Transregio~7 ``Gravitational Wave Astronomy''.  The computations
were performed at the AEI, on the LONI network
(\texttt{www.loni.org}), at LRZ Munich, and the Teragrid (allocation
TG-MCA02N014).

\appendix
\section{Sensitivity curves}
\label{sec:appendix_a}

For convenience, we report below the sensitivity curves used to
compute the SNRs that are often difficult to collect from the
literature.  For LISA we we use the same noise curve as for the LISA
Mock Data Challenge 3 \cite{Babak:2008sn} as implemented by Trias and
Sintes, and made available by the LISA Parameter Estimation Task Force
\cite{LISAPETaskForce}. The noise curve for advanced Virgo can be
found in tabulated form in Ref.~\cite{AdVirgo}.

\begin{widetext}
\begin{eqnarray}
\text{LIGO} \nonumber \\
&\hskip -1.cm
S_h(f) = S_0 \left\{ \left( \frac {4.49 f}{f_0} \right)^{-56} + 
             0.16 \left( \frac{f}{f_0} \right)^{-4.52} + 0.52 + 
             0.32 \left( \frac {f}{f_0} \right)^2 \right\}\,,  
& \hskip -1.cm S_0 = 9 \times 10^{-46}, \quad  f_0 = 150\ {\rm Hz}\,, \nonumber \\
\text{AdLIGO} \nonumber \\
&\hskip -1.cm
S_h(f) = S_0\left\{  \left(\frac{f}{f_0}\right)^{-4.14} -
	5\left(\frac{f_0}{f}\right)^2 + 
	111  \left(1 - \left(\frac{f}{f_0}\right)^2 + 
        \frac{1}{2} \left(\frac{f}{f_0}\right)^4\right)
	\left({1 + \frac{1}{2}\left(\frac{f}{f_0}\right)^2} \right)^{-1}\right\}\,,
&  S_0 = 10^{-49}, \quad f_0 = 215\ {\rm Hz}\,, \nonumber \\
\text{Virgo} \nonumber \\
&S_h(f) = S_0 \left\{ \left( \frac {7.87f}{f_0} \right)^{-4.8} 
	+ \frac{6}{17} \left(\frac{f_0}{f}\right)
    + \left[1 + \left(\frac {f}{f_0} \right)^2 \right] \right\}\,,
& \hskip -1.0cm S_0 = 10.2\times 10^{-46}, \quad f_0 = 500\ {\rm Hz}\,. \nonumber \\
\end{eqnarray}
\end{widetext}




\section{Comparison of waveforms at future null infinity}
\label{sec:appendix_b}

A systematic source of error in the results given in this paper is the
finite radius $r_{_{\rm E}} = 160\,M$ at which our waveforms are
computed. In order to determine its influence on the accuracy of the
values reported here, we have exploited the recent possibility of
computing waveforms unambiguously at future null infinity
$\scri$~\cite{Reisswig:2009us}. In this approach, which
makes use of the Cauchy-characteristic extraction technique
\cite{Bishop93, Bishop96, Bishop98b, Babiuc:2005pg, Babiuc:2009}, the
gravitational-wave information $\Psi_4$ is computed at $\scri$ in a
gauge invariant way and with no causal influence from the outer
boundary.

In practice, we have computed the match between the waveforms
extracted at $r_{_{\rm E}}$ and at $\scri$ for the nonspinning
configuration $s_0$, and found that $\mathcal{M}_{\rm best}=0.999$,
which is thus within the error given by the match between different
numerical resolutions (\cf discussion in Sect.~\ref{sec:match} and see
also Table~\ref{tablethree}). Note that the initial separation of the
two black holes as reported in~\cite{Reisswig:2009us},
$d=11\,M$, is larger than the one reported here, thus resulting in a
much smaller initial frequency $\omega_{\rm ini}$. Nevertheless, we
have considered the same glueing frequency $\omega_{\rm glue}=0.168/M$
so as to have a fair comparison between the two waveforms.

In addition, we have also compared SNRs obtained in the two cases,
when the Fourier-transform of $h(t)$ as given in terms of $\Psi_4$ is
easily obtained as
\begin{equation}
 \tilde{h}(f)=-\frac{\widetilde{\Psi}_4}{4\pi^2 f^2}\,,
\end{equation}
where ${\widetilde \Psi}_4$ is the Fourier-transform of $\Psi_4$. For
any of the total masses considered here and for all of the detectors,
we find that the differences in the SNRs is less than
$1.0\%$. Overall, both results show that the error introduced by the
use of a finite radius calculation is within our numerical error-bars
of $\sim 2.0\%$ and thus does not modify significantly the
results obtained in this work.

\bibliographystyle{apsrev-nourl}
\bibliography{aeireferences}

\end{document}